%% file: main.tex
\documentclass{article}

\usepackage{arxiv}

\usepackage[utf8]{inputenc} % allow utf-8 input
\usepackage[T1]{fontenc}    % use 8-bit T1 fonts
\usepackage{hyperref}       % hyperlinks
\usepackage{url}            % simple URL typesetting
\usepackage{booktabs}       % professional-quality tables
\usepackage{amsfonts}       % blackboard math symbols
\usepackage{nicefrac}       % compact symbols for 1/2, etc.
\usepackage{microtype}      % microtypography
\usepackage[tbtags]{amsmath}
\usepackage{cleveref}       % smart cross-referencing
\usepackage{graphicx}
\usepackage[numbers]{natbib}
\usepackage{doi}
\usepackage{comment}
\usepackage{todonotes}
\usepackage{caption}
\usepackage{subcaption}
\usepackage{comment}
\usepackage{bm}
\usepackage{rotating}

\title{Evaluating the impact of noise on the performance of the Variational Quantum Eigensolver}

\author{Marita Oliv \\
	Fraunhofer Institute for Cognitive Systems IKS\\
	Munich, Germany\\
	\texttt{marita.oliv@iks.fraunhofer.de} \\
	\And
	Andrea Matic\\
	Fraunhofer Institute for Cognitive Systems IKS\\
	Munich, Germany\\
	\texttt{andrea.matic@iks.fraunhofer.de} \\
	\And
	Thomas Messerer\\
	Fraunhofer Institute for Cognitive Systems IKS\\
	Munich, Germany\\
	\texttt{thomas.messerer@iks.fraunhofer.de} \\
	\And
	Jeanette M. Lorenz\\
	Fraunhofer Institute for Cognitive Systems IKS\\
	Munich, Germany\\
	\texttt{jeanette.miriam.lorenz@iks.fraunhofer.de} \\	
}

\renewcommand{\headeright}{}
\renewcommand{\undertitle}{A preprint}
\renewcommand{\shorttitle}{Evaluating the impact of noise on the performance of the Variational Quantum Eigensolver}

\hypersetup{
pdftitle={Evaluating the impact of noise on the performance of the Variational Quantum Eigensolver},
pdfsubject={quant-ph},
pdfauthor={M. Oliv, A. Matic, T. Messerer, J.Lorenz},
pdfkeywords={Variational Quantum Eigensolver, Noise, Quantum chemistry , Quantum Simulation , Variational Algorithms},
}

\begin{document}
\maketitle

\begin{abstract}
	Quantum computers are expected to be highly beneficial for chemistry simulations, promising significant improvements in accuracy and speed. The most prominent algorithm for chemistry simulations on NISQ devices is the Variational Quantum Eigensolver (VQE). It is a hybrid quantum-classical algorithm which calculates the ground state energy of a Hamiltonian based on parametrized quantum circuits, while a classical optimizer is used to find optimal parameter values. However, quantum hardware is affected by noise, and it needs to be understood to which extent it can degrade the performance of the VQE algorithm. In this paper, we study the impact of noise on the example of the hydrogen molecule. First, we compare the VQE performance for a set of various optimizers, from which we find NFT to be the most suitable one. Next, we quantify the effect of different noise sources by systematically increasing their strength. The noise intensity is varied around values common to superconducting devices of IBM Q, and curve fitting is used to model the relationship between the obtained energy values and the noise magnitude. Since the amount of noise in a circuit highly depends on its architecture, we perform our studies for different ansatzes, including both hardware-efficient and chemistry-inspired ones. 
\end{abstract}


\section{Introduction}

\input{introduction}

\section{Related work}

\input{related_work}

\section{Background}

\input{Background}

\section{Experimental setup and results}

\input{experiment}

\section{Conclusion}

\input{conclusion}

\section*{Acknowledgements}
The research is part of the Munich Quantum Valley, which is supported by the Bavarian state government with funds from the Hightech Agenda Bayern Plus.

\appendix
\input{appendix}

\newpage

\bibliographystyle{ieeetr}
\bibliography{main.bib}

\end{document}

%% file: introduction.tex
Quantum chemistry constitutes one of the most promising fields for the application of quantum computers~\cite{preskill_quantum_2018, cao_quantum_2019, mcardle_quantum_2020, outeiral_prospects_2021}. For this purpose, a variety of quantum algorithms has been proposed~\cite{motta_emerging_2022}. Due to the lack of fault-tolerant quantum computers, a major focus of current research lies on hybrid quantum-classical algorithms, which can be executed on Noisy Intermediate Scale Quantum (NISQ) devices. A prominent example is the Variational Quantum Eigensolver (VQE)~\cite{peruzzo_variational_2014}. Initially proposed to calculate the ground state energy of molecules, this algorithm allows e.g. to predict chemical reaction rates or to find optimal molecular geometries. Apart from quantum chemistry, VQE can be used for a wide range of problems in other fields, including condensed matter physics and optimization~\cite{cao_quantum_2019, tilly_variational_2022, moll_quantum_2018, mohanty_analysis_2022}. 

The VQE algorithm is designed to find the ground state energy of a Hamiltonian by calculating its expectation value in a parametrized state $\psi(\bm{\theta})$. According to the variational principle, this expectation value represents an upper bound on the true ground state energy. To minimize the expectation value, the parameters $\bm{\theta}$ are iteratively optimized using a classical computer until the algorithm converges.
 To obtain a successful result it is essential to choose a good ansatz for $\psi(\bm{\theta})$, i.e. a parametrized quantum circuit with a suitable architecture. One generally distinguishes between hardware-efficient ansatzes, which are easy to implement on hardware, and chemistry-inspired ansatzes, where the circuit construction is based on prior chemistry knowledge. Apart from the ansatz, also the classical optimizer has an impact on the quality of the results. Depending on the optimizer and the specific problem settings, the optimization process can be hindered from reaching the global minimum of the loss function by converging into local minima and getting stuck on barren plateaus. Therefore, the optimizer needs to be chosen carefully. 

Quantum hardware devices are always subject to quantum noise. It arises in all steps of the quantum circuit, from the state preparation over the applied gates to the measurement. Thus, noise also influences the optimization process, where it has been shown to introduce additional traps to the loss landscape \cite{wang_noise-induced_2021}. Therefore, the VQE performance for a particular optimizer may substantially differ between noiseless simulations and experiments with noise. Furthermore, as the total amount of noise in a hardware device depends on the number and types of gates in the quantum circuits, the optimization process and thus the suitability of an optimizer may vary between different ansatze. 

There are several approaches to mitigate the effect of noise for the VQE algorithm~\cite{tilly_variational_2022}. However, the extent of its noise susceptibility is not yet fully understood. Because of that it is unclear which error mitigation method should be chosen for a certain problem and which requirements the hardware needs to fulfill to allow for a successful result. 
In this paper we therefore investigate the effect of noise on VQE in a systematic way. We perform the studies based on the example of calculating the ground state energy of the hydrogen molecule H$_2$, which is computationally easy to simulate due to its small size. Additionally, it has the advantage of precisely known energy levels. First, we compare the performance of different classical optimizers to find a suitable one for the subsequent studies. After that, we investigate how the obtained ground state energy changes in the presence of noise. 
To get a detailed understanding of the underlying effects, we model each noise type separately in simulations and calculate the energies for varying noise strengths. 

The paper is structured as follows: Section~\ref{sec:related_work} gives an overview of the related work. In section~\ref{sec:background} the VQE algorithm is explained and the most prevalent types of quantum noise are discussed. Section~\ref{sec:experiment} describes our experiments and their results. Finally, in section~\ref{sec:conclusion} we summarize our results and give an outlook on possible future research directions.

%% file: related_work.tex
\label{sec:related_work}
Quantifying the influence of noise on the performance of VQE requires a thorough analysis on its different components. The works by \cite{saib_effect_2021} and \cite{sennane_calculating_2022} studied the connection between noise and the ansatz choice: In \cite{saib_effect_2021} the H$_2$ molecule is simulated using different hardware-efficient ansatzes. It was found that the ranking of the best performing circuits significantly changes when including noise effects. In \cite{sennane_calculating_2022} the performance of a hardware-efficient and a chemistry-inspired ansatz was compared by simulating the ground state energy of benzene. Since the chemistry-inspired ansatz has deeper circuits, the accuracy of the results is strongly affected by noise. With the hardware-efficient ansatz it was possible to outperform classical simulation approaches for strong molecule deformations. However, the results were also shown to be sensitive to the parameter initialization. 
A comprehensive analysis on the accuracy of VQE is presented in \cite{mihalikova_cost_2022}, where the ground state energy of H$_2$ was simulated for different variations of the algorithm settings. This includes e.g. a systematic variation of the number of shots, a comparison of different hardware-efficient ansatzes as well as the evaluation of the results with and without noise, for which the noise model of one particular IBM Quantum device is used.  
In addition to the total hardware noise, readout and gate errors were also studied separately in the simulations. 

These studies have in common that the noise was applied with a fixed intensity. We extend these studies by investigating how the results of the VQE algorithm change when the different noise components are gradually increased. This approach was used in \cite{lee_error-mitigated_2021}, where the positively charged helium hydride ion was simulated for a photonic hardware system. For this, the obtained energies were compared for increasing intensities of depolarizing noise. Similarly, different strengths of depolarizing and readout noise were tested in \cite{zhang_comparative_2022}. Here, the ground state energy of the H$_2$ molecule was studied as a function of the interatomic distance. However, the focus of both these studies lies on error mitigation techniques rather than on a detailed noise modelling.

Apart from molecular simulations, comprehensive noise studies can be found in the works by \cite{zeng_simulating_2021, fontana_evaluating_2021, cui_optimization_2021}. In \cite{zeng_simulating_2021} the impact of noise was analyzed for the ground state energy of different lattice spin models. The VQE simulations included amplitude, phase damping and depolarizing noise with different strengths. These noise sources were also investigated in \cite{fontana_evaluating_2021}, where the ground state energy of a transverse-field Ising Hamiltonian was calculated. It was shown that quantum circuits become more resilient to noise if redundant parametrized gates are added. In \cite{cui_optimization_2021} the impact of different noise sources was analyzed for the problem of solving the one-dimensional Poisson equation. Instead of VQE, a custom algorithm was used in these studies. 

The suitability of different classical optimizers for noisy VQE simulations was investigated in the works by \cite{mihalikova_best-practice_2022, wright_numerical_nodate}. In \cite{mihalikova_best-practice_2022} the accuracy of the H$_2$ simulations was compared for the four optimizers Simultaneous Perturbation Stochastic Approximation (SPSA), COBYLA, Nelder-Mead and Powell. The most accurate results were achieved with SPSA and COBYLA. Additionally, SPSA was found to be relatively robust against varying the choice of entangling gates in the circuit. In \cite{wright_numerical_nodate} the ground state energy of sodium hydride was calculated using the COBYLA and L-BFGS optimizers. Different chemistry-inspired ansatzes were tested and the accuracy was evaluated for varying strengths of depolarizing noise. For all these experiments both optimizers lead to similar energy values. 

The studies in \cite{lavrijsen_classical_2020} provided an extensive comparison of different optimizers. Considering the problem of simulating molecular processes in ethylene, noise effects were implemented with a Gaussian-based model and the optimization results were evaluated for varying noise strengths. In the case of zero noise, optimizers of the SciPy package \cite{virtanen_scipy_2020}, like BFGS or COBYLA, were the fastest ones. However, they lead to insufficent results in the presence of noise. Therefore, the authors recommend to use other optimizers. They also found that the optimizer choice is highly connected to the problem statement, e.g. whether good parameter bonds are available beforehand. 

In \cite{nakanishi_sequential_2020} a sequential optimization method (NFT) was introduced and tested for VQE simulations, e.g. for lithium hydride. The NFT optimizer was found to converge better and faster compared to other optimizers such as SPSA or BFGS, even when a large statistical error was applied. This indicates that NFT may also be robust against quantum noise. To improve the accuracy of VQE in the presence of noise, \cite{iannelli_noisy_2021} used a Bayesian optimization technique based on Gaussian process regression. For the studied problem of a transverse-field Ising model the Bayesian optimizer clearly outperformed SPSA and showed a similar performance as NFT. 

Except for \cite{mihalikova_best-practice_2022}, the presented optimizers were tested for other problems than the ground state energy calculation of hydrogen. In our studies, we thus build up upon \cite{mihalikova_best-practice_2022} and investigate the performance for hydrogen using additional optimizers, such as NFT and the Bayesian optimizer.

%% file: Background.tex
\label{sec:background}
\subsection{Principle of VQE}
The Variational Quantum Eigensolver (VQE) \cite{peruzzo_variational_2014, tilly_variational_2022} is a quantum-classical hybrid computational method to find the smallest eigenvalue of a Hamiltonian $H$, which describes the energy of a physical system like a molecule or a nucleus. It is based on the Rayleigh-Ritz variational principle, which states that the ground state energy $E_0$ can be bounded by

\begin{equation} 
    \label{rayleigh-ritz}
    E_0 \le \min _{\bm\theta} \langle \psi(\bm\theta) | H | \psi(\bm\theta) \rangle,
\end{equation}

where $| \psi(\bm\theta) \rangle$ is an appropriate parameterized state. 
%The VQE implements this principle as hybrid algorithm with an iterative optimization of the parameters $\bm\theta$.
Using the Born-Oppenheimer approximation \cite{born_zur_1927}, the Hamiltonian of a molecule is represented in second quantization by 

\begin{equation} \label{secondQuant H}
    H= \sum_{pq} h_{pq} a_p^\dagger a_q + \frac{1}{2} \sum_{pqrs} h_{pqrs} a_p^\dagger a_q^\dagger a_r a_s ,
\end{equation}

with the fermionic annihilation and creation operators $a$ and $a^\dagger$. The one- and two-body integrals $h_{pq}$ and $h_{pqrs}$ describe the kinetic energy of the electrons, the Coulomb interaction between nucleus and electrons as well as the Coulomb interaction between the electrons.
%\begin{gather}
%    h_{pq} = \int \Phi_p^*(r) \left( - \frac{1}{2} \nabla ^2 - \sum_I \frac{Z_I}{R_I -r}\right) \Phi_q(r) dr \\
%    h_{pqrs}= \int \frac{\Phi_p^*(r_1)\Phi_q^*(r_2)\Phi_r(r_2)\Phi_s(r_1)}{|r_1-r_2|}dr_1dr_2,
%\end{gather} where $\Phi_i$ is the wavefunctions of the single electron i and $r_i$ denotes the respective spatial position. 
While $h_{pq}$ and $h_{pqrs}$ are calculated classically, the fermionic operators $a$ and $a^\dagger$ need to be mapped onto operations which can be executed by a quantum computer. There are several possible mappings available for this purpose. A commonly used one is the Jordon-Wigner mapping \cite{jordan_uber_1928}, where $a$ and $a^\dagger$ are translated to tensor products of Pauli operators $Z$, $Y$ and $X$ that can be implemented as gates on a quantum computer. With this mapping the Hamiltonian can be expressed as a linear combination of Pauli strings $P_k$ -- tensor products of Pauli operators -- with the classically calculable, real coefficients $h_k$

\begin{equation} 
    \label{H as Paulistrings}
    H = \sum_k h_k P_k .
\end{equation}

Therefore, the variational principle in \ref{rayleigh-ritz} can be reformulated as

\begin{equation}
    E_0 \le \min _{\bm\theta} \left( \sum_k h_k \langle  \psi(\bm\theta) | P_k | \psi(\bm\theta) \rangle \right).
\end{equation} 

In the VQE algorithm the expectation values of each of the Pauli strings are computed on a quantum device, while a classical computer is used to calculate their linear combination and to adjust the parameters $\bm\theta$ such that the expectation value of $H$ is minimized. 
The parameter optimization and the calculation of the expectation value are repeated iteratively until the algorithm converges. 
%This hybrid structure has the great advantage of an efficient calculation of the cost function with short quantum circuits with a few tens to hundreds gates \todo{Zahlen mit Quellen belegen}. For those no long coherence times are needed, which makes the VQE appropriate for NISQ devices.

The parametrized state $| \psi(\bm\theta) \rangle$, called ansatz, is represented as a parametrized unitary $U(\bm\theta)$ acting on a fixed reference state $| \psi_0 \rangle$.
%\begin{equation}
%    | \psi(\bm\theta) \rangle = U(\bm\theta) | \psi_0 \rangle .
%\end{equation} 
A common choice for $|\psi_0 \rangle$ is the Hartree-Fock wave function. In the Hartree-Fock or mean-field theory, the 
%$n$-body problem is separated into $n$ one-body problems, where the 
total wave function is given by a linear combination of the single wave functions. This makes it classically calculable, but neglects electron correlations. To yield a more refined approximation of the exact ground state wave function, a suitable unitary needs to be chosen which accounts for these neglected effects. 

As for the Hamiltonian, the wave function must be mapped onto the quantum computer. A possibility for this is to map each spin orbital, which can be either empty or occupied by an electron, onto a qubit, which is initialized in the state $|0\rangle$ or $|1\rangle$. The unitary operation for the ansatz can be hardware-efficient or chemically-inspired. While for a hardware-efficient ansatz practical considerations on the implementation of the quantum computer determine the design, a chemically-inspired ansatz seeks to model the neglected effects using prior chemical knowledge about the system and its symmetries. Typically, this leads to much longer quantum circuits. 
%There is a trade-off between the possible advantages of symmetries and of short circuits, which in general suffer less of noise from the quantum device, for the accuracy of the solution. \todo{MO: for the accuracies.. am ende klingt etwas merkwürdig}

An example for a chemically-inspired ansatz is a unitary coupled-cluster ansatz based on single and double excitations (UCCSD)~\cite{peruzzo_variational_2014, romero_strategies_2017}.
It originates from the classical coupled-cluster theory which is modified to be unitary. This leads to the unitary coupled-cluster (UCC) approach

\begin{equation} \label{ucc equ}
| \psi \rangle = e^{T-T^\dagger}|\psi_0\rangle ,
\end{equation} 

where the excitation operator $T$ is defined as 

\begin{equation}
T= \sum _i T_i .\\
%T_1 = \sum _{i \in occ, a \in virt} t_a ^i \tau_a ^\dagger \tau_i \\
%T_2 = \sum _{i>j \in occ, a>b \in virt} t_{ab} ^{ij} \tau_a ^\dagger \tau_b^\dagger \tau_i \tau_j \\
%...
\end{equation}

$T_i$ represents the excitation of $i$ electrons from occupied orbitals of the reference state to unoccupied sites. For the UCCSD ansatz $T$ is truncated to consider only single and double excitations $T_1$ and $T_2$, respectively. As H$_2$ has only two electrons, the UCC and UCCSD ansatzes are equivalent in this case.

\subsection{Noise} \label{noise theory}

There are two types of noise in the evolution of a quantum state: coherent and incoherent noise. %Coherent noise is reversible and can thus be represented as unitary transformation. Miscalibrated gates are an example for this.
Coherent noise is reversible and is caused e.g. by miscalibrated gates.
Incoherent noise, on the other hand, arises due to interactions with the environment. 
%Therefore, the qubits in a quantum computer consitute an open system, which is in general not described by unitary transformations. Only together with their environment they can be considered as a closed system.
Mathematically, the principal system and the environment are each described by a density matrix, $\rho$ and $\rho_{env}$. Their interaction is represented by a unitary and one is interested in the state of the system afterwards $\mathcal{E}(\rho)$.
The map $\rho \to \mathcal{E}(\rho)$ is a quantum channel and can be described in the operator-sum representation by~\cite{nielsen_quantum_2010}

\begin{equation}
    \mathcal{E}(\rho)= \sum_k E_k \rho E_k^\dagger ,
\end{equation}

with the Kraus operators $E_k$ which fulfill $\sum_k E_k^\dagger E_k \le I$ in order to allow for the interpretation of $tr(E_k \rho E_k^\dagger)$ as probability of an outcome $k$ from a measurement. Different types of noise are modeled by different Kraus operators as described in the following.

\subsubsection{Readout noise}
Measurements result in a binary outcome.
With probability $p$ the measurement fails, giving the opposite state of the intended one. This noise channel can thus be described as bit-flip channel with Kraus operators \begin{equation}
    E_0=\sqrt{1-p}I , \quad E_1=\sqrt{p}X ,
\end{equation} where $I$ is the identity matrix.

\subsubsection{Depolarizing noise}
Random fluctuations of e.g. electromagnetic fields in the environment can cause a qubit to depolarize with probability $p$, i.e. its state is replaced by the completely mixed state $I/2$. So, after the interaction with the environment the state of the qubit is given by $\mathcal{E}(\rho)= (1-p) \rho + pI/2$ corresponding to Kraus operators \begin{equation}
    E_0=\sqrt{1-p}I, \quad E_1=\sqrt{\frac{p}{3}}X, \quad E_2=\sqrt{\frac{p}{3}}Y, \quad E_3=\sqrt{\frac{p}{3}}Z.
\end{equation} 

It corresponds to a combined bit-flip ($X$), phase-flip ($Z$) and bit-phase flip channel ($Y=iXZ$).

\subsubsection{Amplitude and phase damping noise}
Amplitude damping noise describes energy dissipation to the environment.
The quantum channel can be represented by the Kraus operators \begin{equation}
    E_0=\sqrt{1-\epsilon} \left[ \begin{array}{cc} 1&0 \\0 & \sqrt{1-p_a} \end{array}\right], \quad
E_1=\sqrt{1-\epsilon} \left[ \begin{array}{cc} 0& \sqrt{p_a} \\0 & 0 \end{array}\right], \quad
 E_2=\sqrt{\epsilon} \left[ \begin{array}{cc} \sqrt{1-p_a}&0 \\0 & 1 \end{array}\right], \quad
 E_3=\sqrt{\epsilon} \left[ \begin{array}{cc} 0& 0 \\\sqrt{p_a} & 0 \end{array}\right] ,
\nonumber
\end{equation}

where $p_a$ is the probability for amplitude damping and $\epsilon$ determines the probability of excitation in case of an error, so that after infinitely many interactions with the environment the qubit is in the equilibrium state $$\rho_{\infty}=\left[ \begin{array}{cc} \epsilon& 0 \\\ 0 & 1-\epsilon \end{array}\right].$$ 

%Here, $\epsilon$ depends on the temperature of the environment as an approximate Maxwell-Boltzmann statistics \cite{jin_thermal_2015}. 
Taking into account the temperature of superconducting qubits of $\sim15$~mK and an average qubit frequency of the order of $10^9$~Hz~\cite{ibmq}, the probability for an excitation $\epsilon$ is in the order of $10^{-14}$~\cite{jin_thermal_2015}. It is therefore a reasonable approximation to set $\epsilon=0$. Thereby the set of Kraus operators is consequently reduced to $E_0$ and $E_1$.
The amplitude damping channel acts on both the diagonal and off-diagonal elements of the density matrix, driving it towards the equilibrium state. 
%The $E_1$ operator has the effect of changing the state $|1\rangle$ into $|0\rangle$, corresponding to the physical process of losing a quantum of energy to the environment. $E_0$ leaves a state $|0\rangle$ unchanged, but reduces the amplitude of a $|1\rangle$ state; this can be interpreted physically as the environment perceiving it to be more likely that the system is in the $|0\rangle$ state, rather than the $|1\rangle$ state, because no energy was lost.

Phase damping noise describes the loss of quantum information without the loss of energy, in other words, the transition towards classical behavior. In the operator-sum framework a possible representation is by the Kraus operators \begin{equation}
    E_0=\left[ \begin{array}{cc} \ 1 & 0 \\\ 0 & \sqrt{1-p_\phi} \end{array}\right], \quad
    E_1=\left[ \begin{array}{cc} 0& 0 \\\ 0 & \sqrt{p_\phi} \end{array}\right] .
\end{equation} It has the effect of diminishing the off-diagonal elements of the initial density matrix proportional to the probability $p_\phi$.

%Another possible representation is given by the Kraus operators (8.130) \begin{equation}
 %   \tilde E_0= \sqrt{\kappa}I \quad \tilde E_1= \sqrt{1- \kappa} Z
%\end{equation} with $\kappa=(1+\sqrt{1-p_p})/2$ leading to the interpretation of the phase damping as a phase flip occurring with probability $\kappa$.

For most cases, it is sensible to define the probability for amplitude or phase damping to be time-dependent. A common choice, motivated by physical experiments, is an exponential decay
\begin{equation}
    p_a=1-e^{-t/T_1}, \quad p_\phi=1-e^{-t/2 T_\phi}
\end{equation}

with characteristic times $T_1$ and $T_\phi$. Since amplitude and phase damping both affect the off-diagonal elements, they are commonly combined in the relaxation time $T_2$:

$$\frac{1}{T_2}=\frac{1}{2T_1}+\frac{1}{T_\phi}$$ .

Amplitude and phase damping together make up thermal relaxation noise.

\subsubsection{Shot noise}
A conceptually different kind of noise is stochastic or shot noise. To reconstruct the probability distribution corresponding to a state prior to the measurement, it is approximated by a finite number of measurements $N$ of the state. According to the central limit theorem and the properties of Gaussian distributions, the uncertainty on a measured expectation value $\sigma$ is proportional to $1/\sqrt{N}$.

%% file: experiment.tex
\label{sec:experiment}
\subsection{Experimental setup}

To investigate the effect of noise on the performance of VQE, we simulate the hydrogen molecule H$_2$ using the Pennylane framework \cite{bergholm_pennylane_2018}. For simulating noise, the models of Qiskit Aer \cite{Qiskit} are used. These simulations replicate the process of obtaining probabilities by a finite number of measurements and thus always include shot noise.
To obtain a more detailed insight into the significance and effect of noise, different kinds of noise are simulated separately. In particular, we assess the results of simulations with shot noise only as well as with readout, depolarizing, amplitude and phase damping noise included. These are compared to a statevector simulation which calculates the resulting probability distributions analytically and thus is not subject to any kind of noise.
The models implement noise as additional operations on the qubits. Depolarizing, amplitude and phase damping noise are applied after each gate, readout noise before each measurement. 
The same error rate is chosen for all qubits and gates, except for depolarizing noise, where we distinguish between 1-qubit and 2-qubit gates. This choice is made to reduce the number of variables of the noise models to a reasonable size while still accounting for the fact that on superconducting processors the error rates for 2-qubit gates are typically at least 10 times higher than for 1-qubit gates~\cite{ibmq}. For an estimation of the relevant noise intensities the typical properties, namely the gate lengths, relaxation times and error rates of the fundamental gates, of the IBM Quantum Falcon r5.11 version 1 processors are considered.

\graphicspath{{./imagesSetup/}}

\begin{figure} 
    \centering
    \includegraphics[width=0.33\textwidth]{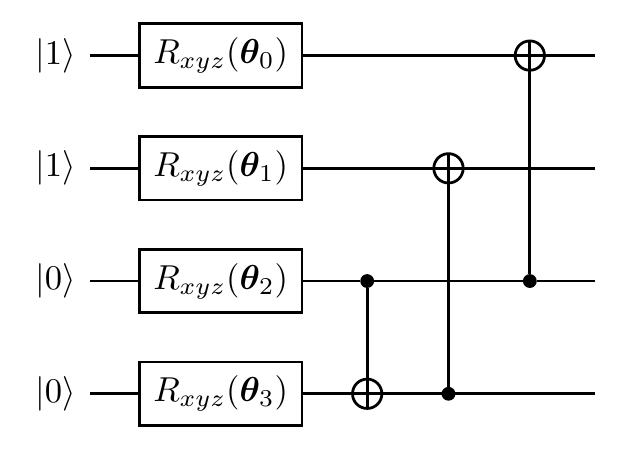}
    \caption{The hardware-efficient $R_{XYZ}$ ansatz used for the simulations.}
    \label{fig:ansatzes}
\end{figure}

To yield more general insights the simulations are done for three different ansatzes, two hardware-efficient and a chemically-inspired one.
The first hardware-efficient ansatz, named $R_{XYZ}$ in the following, is a general one with small depth. Its structure is shown in figure \ref{fig:ansatzes}. The first $2n$ qubits are initialized in state $|1 \rangle$, where $n$ is the number of electrons in the molecule. On each qubit a rotation around all axes is applied, followed by an entangling layer of CNOT gates. Thus, the number of variational parameters for this ansatz is three times the number of qubits, i.e. 12 parameters for our setup.
As example for a chemically-inspired ansatz UCCSD is chosen because of its frequent usage in the literature to facilitate comparability. The number of variational parameters for this ansatz is given by the number of possible one- and two-electron excitations. In the case of H$_2$ with four spin-orbitals, consequently three parameters are used. Its depth of 37 is considerably higher than the one of the $R_{XYZ}$ ansatz.
The third ansatz, denoted by $R_Y$, is chosen to have a number of parameters comparable to that of UCCSD, but with an architecture similar to the $R_{XYZ}$ ansatz. Therefore, the general rotations in the hardware-efficient $R_{XYZ}$ ansatz are replaced by rotations around the y-axis only. Thus, the number of parameters of this ansatz is equal to the number of qubits, i.e. four parameters for our experiments. The UCCSD and $R_Y$ ansatz are shown in appendix~\ref{app:ansatzes}. 

\subsection{Hydrogen molecule}
For the simulations of the H$_2$ molecule we use the 4-qubit Hamiltonian \cite{bergholm_pennylane_2018}
%\begin{equation} 
\begin{equation} \label{H2Ham}
    %\begin{aligned}
        %\begin{split}
    %\begin{multline}
    \begin{aligned}
        H &=c_1 I_0
        + c_2 Z_0
        + c_3 Z_1 \\
        &+ c_4 \left(Z_2 + Z_3\right) \\
        &+ c_5 \left(Z_0 Z_2+  Z_1 Z_3 \right)
        + c_6 \left(Z_0 Z_3+  Z_1 Z_2\right)\\
        &+ c_7 Z_0 Z_1
        + c_8 Z_2 Z_3 \\
        &+ c_9 \left( Y_0 X_1 X_2 Y_3+ X_0 Y_1 Y_2 X_3 - Y_0 Y_1 X_2 X_3 - X_0 X_1 Y_2 Y_3\right)
    \end{aligned}
   %\end{multline}
        %\end{split}
    %\end{aligned}
\end{equation} 

with prefactors $c_i$ at an interatomic distance of 1.3228 au. The values of $c_i$ are listed in appendix~\ref{appendix:hamiltonian}.
The minimal STO-3g basis set is used to model the molecular orbitals in the Hamiltonian.
The exact, classically calculated ground state energy for this problem is -1.136189454088 Ha~\cite{pennylane_VQE}.
Despite the simplicity of the hardware-efficient ansatzes used, in statevector simulations the VQE is able to find energies with high accuracy up to the order of $10^{-7}$~Ha for all three ansatzes. So, the ansatzes are expressive enough to cover the solution.

\subsubsection{Optimizer studies} \label{Optimizers}
\graphicspath{{./imagesOptimizers/pdf/}}

Before detailed studies on the noise impact can be performed, one needs to ensure that the applied optimizer performs well in the presence of noise. Therefore, this section compares the behavior of different classical optimizers in VQE simulations of the H$_2$ molecule.
We use a selection of different gradient-based as well as gradient-free optimizers for our studies. This includes the optimizers Nelder-Mead \cite{10.1093/comjnl/7.4.308, virtanen_scipy_2020} and SPSA \cite{james_c_spall_implementation_1998, virtanen_scipy_2020}, which are frequently used for VQE problems, e.g. in the works by \cite{peruzzo_variational_2014, kandala_hardware-efficient_2017, mihalikova_best-practice_2022}. Based on the discussions in section~\ref{sec:related_work}, we further take into account NFT\cite{nakanishi_sequential_2020} and the Bayesian optimizer \cite{Kushner1963ANM, Timgates422020Scikit-optimize}, as these are expected to be highly resilient to noise. In addition, we also perform tests with the Adam optimizer \cite{kingma_adam_2017}, which is a widely used optimizer in classical machine learning problems. Furthermore, we introduce a modification of the SPSA algorithm which we call SPSAreopt. It is based on a two-step process of a coarse search followed by a finer one, which showed improved results compared to the standard SPSA in our experiments. More details on each method can be found in appendix \ref{appendix:optimizers}.

\begin{figure}[h!]
    \centering
    \includegraphics[width=0.7\textwidth]{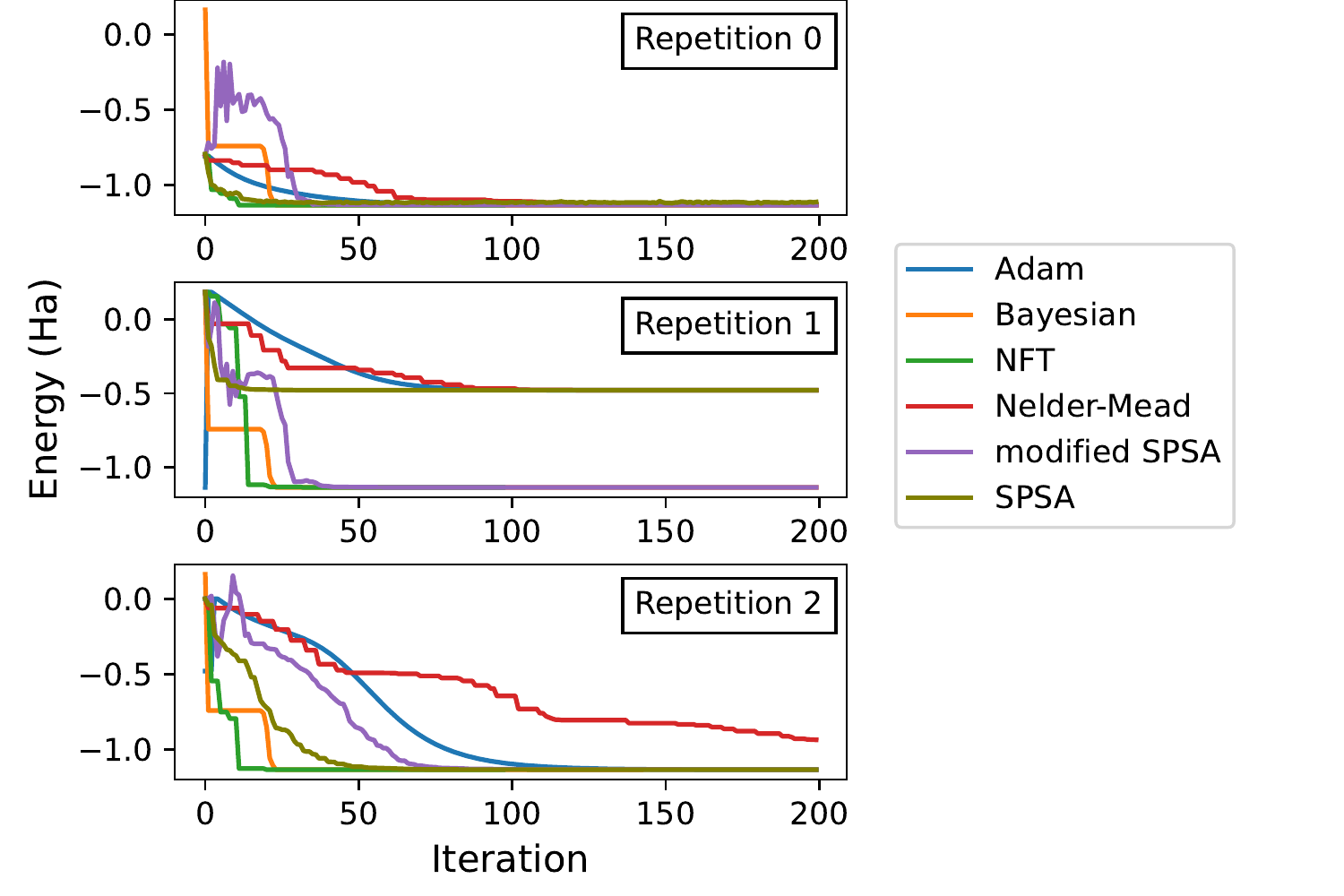}
    \caption{Energy as a function of the number of iteration for various optimizers for three repetitions with different sets of initial parameters. }
    \label{fig:allOptimizers}
\end{figure}

For our experiments the hyperparameters of each optimizer are adjusted such to yield the best possible results in statevector simulations with the hardware-efficient $R_{XYZ}$ ansatz. To cross-validate our results, we perform the studies for five different sets of initial parameters. Figure \ref{fig:allOptimizers} shows the learning curves for the tested optimizers for three sets of initial parameters. While for some initial parameters all optimizers find energies near the ground state energy, for others most optimizers converge into local minima.

\begin{figure}[h!]
    \centering
            \begin{subfigure}[b]{0.33\textwidth}
    \includegraphics[width=\textwidth]{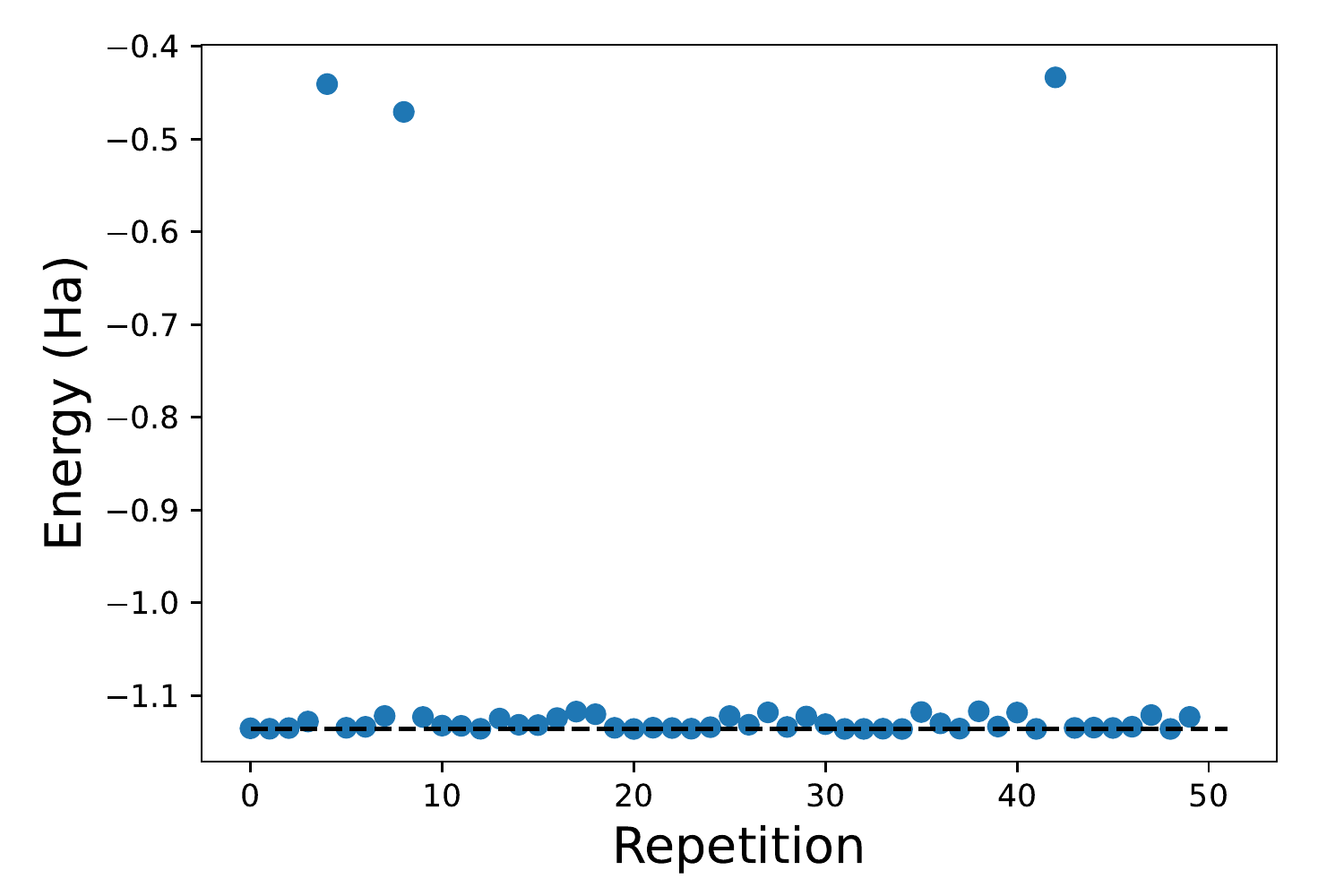}
    \caption{Modified SPSA with $R_{XYZ}$ ansatz}
  \end{subfigure}
         \begin{subfigure}[b]{0.33\textwidth}
    \includegraphics[width=\textwidth]{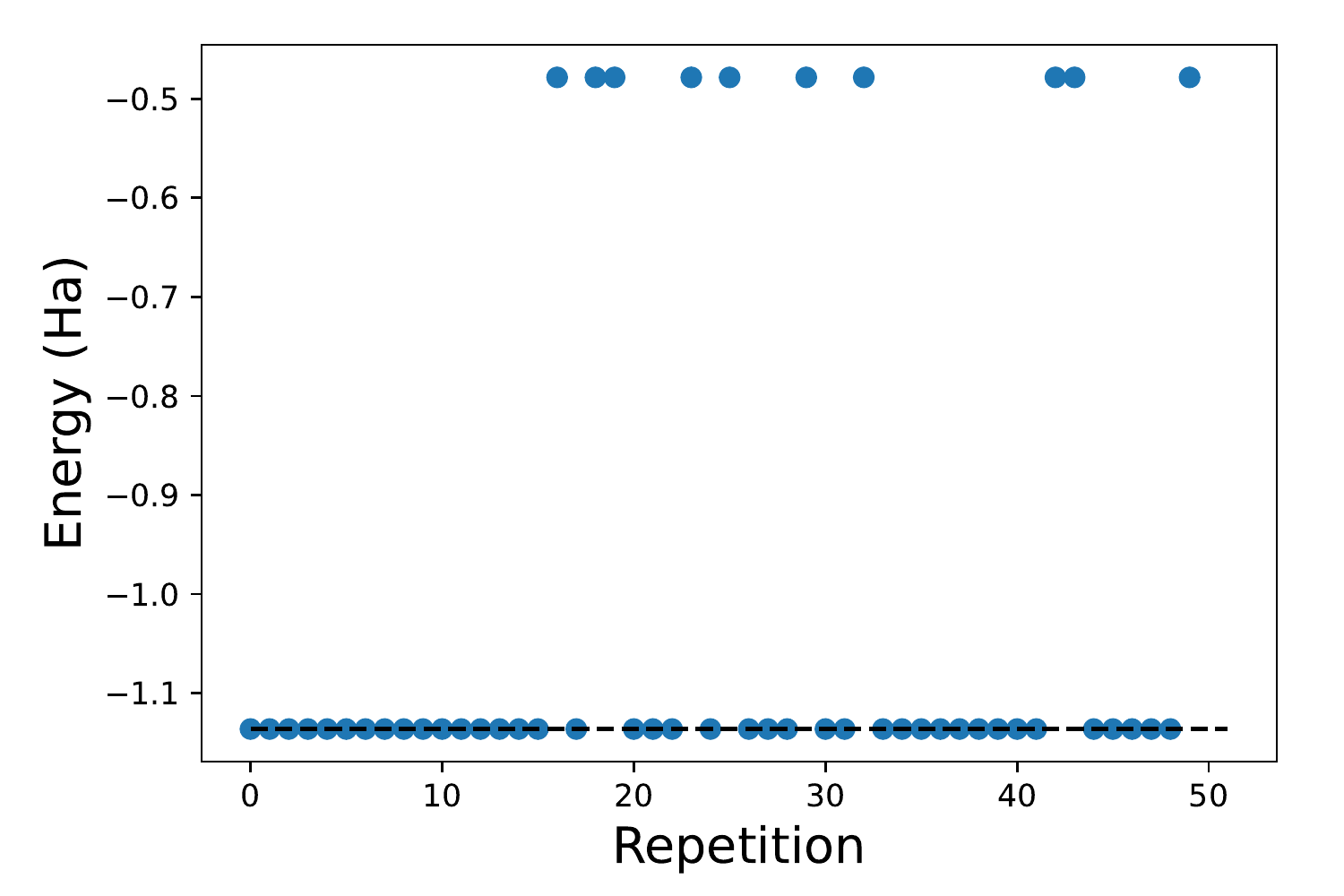} 
    \caption{NFT with $R_{XYZ}$ ansatz}
   \end{subfigure}
          \begin{subfigure}[b]{0.33\textwidth}
    \includegraphics[width=\textwidth]{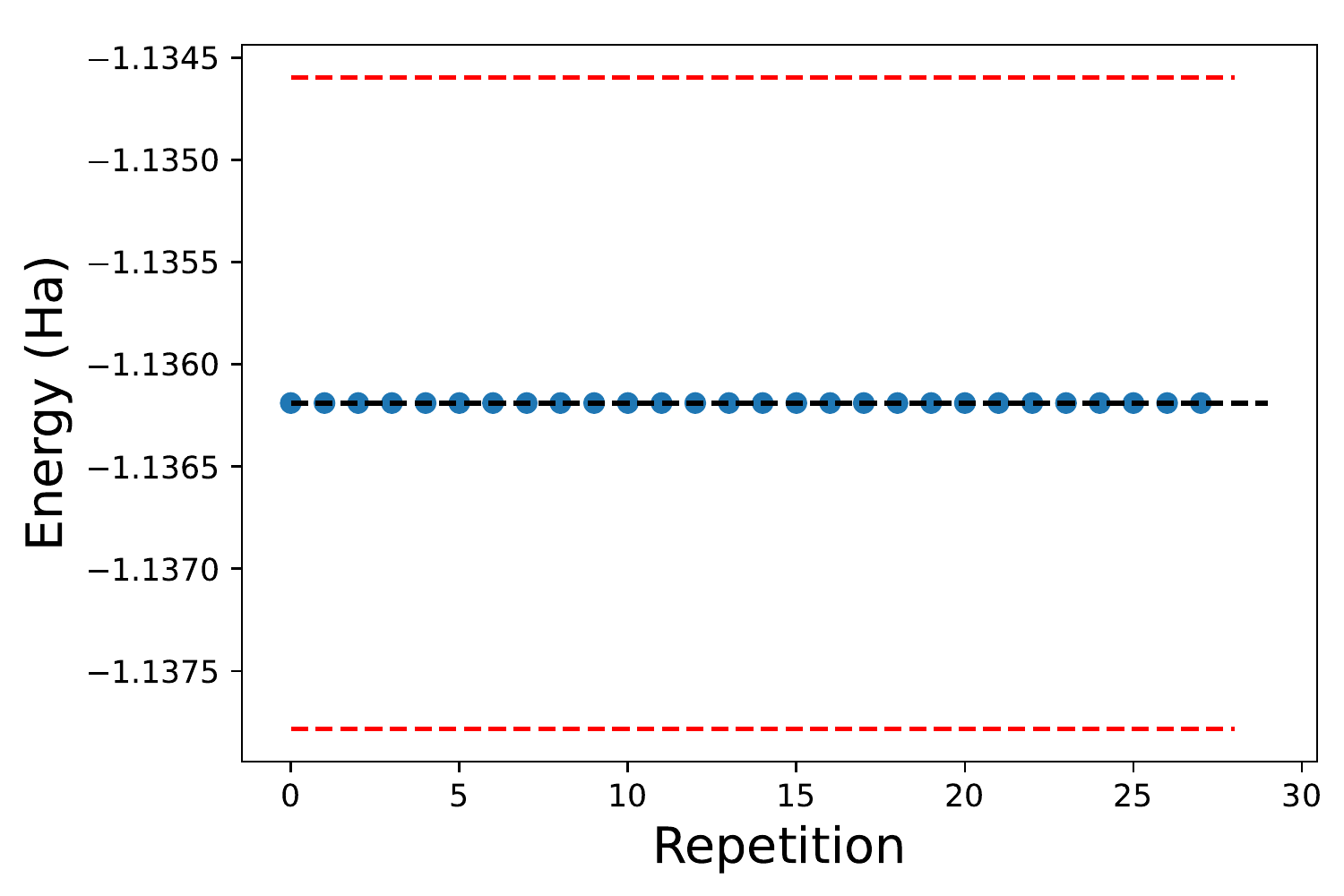}
    \caption{Bayesian with $R_{XYZ}$ ansatz}
  \end{subfigure}
         \begin{subfigure}[b]{0.33\textwidth}
    \includegraphics[width=\textwidth]{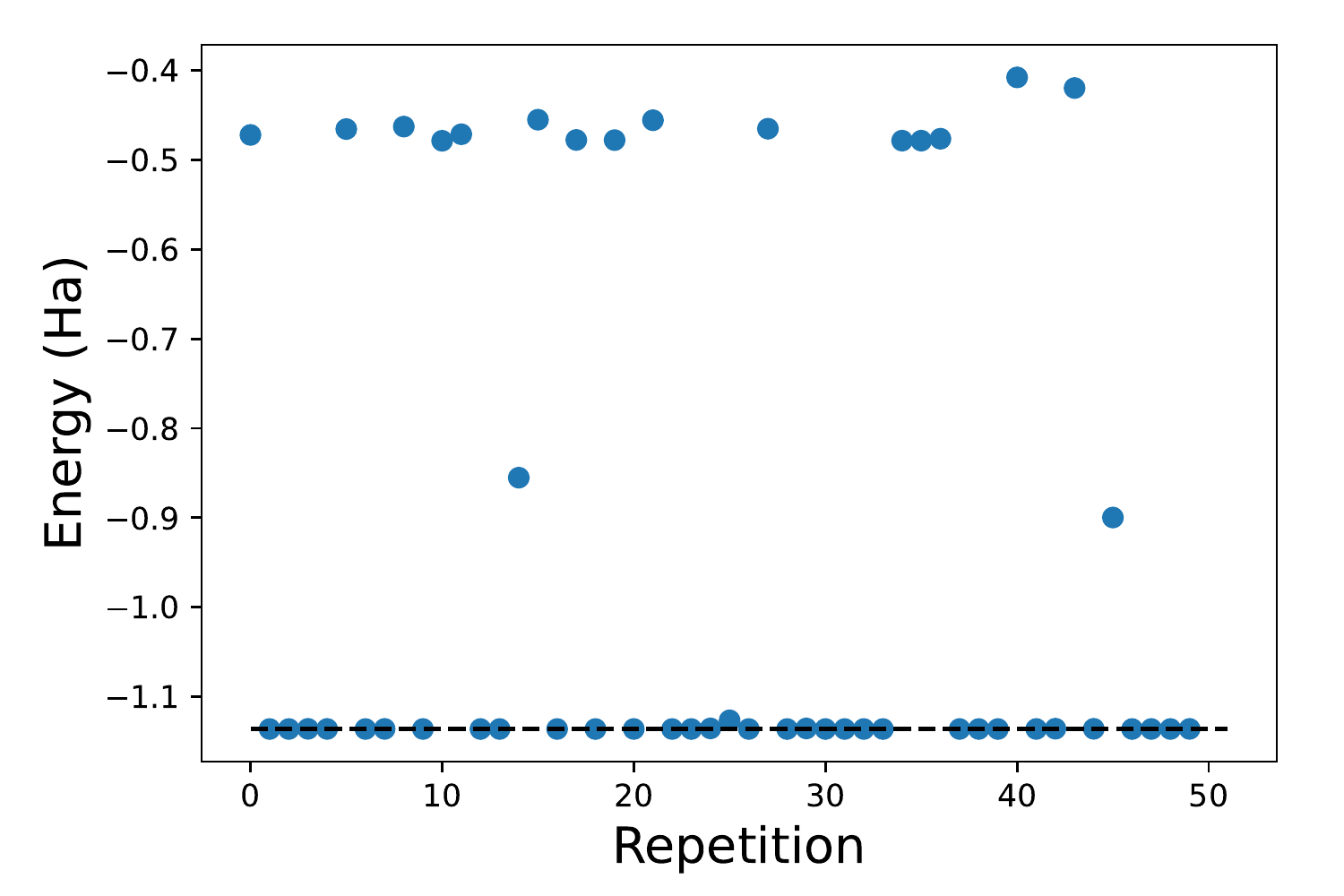}
    \caption{Modified SPSA with $R_{Y}$ ansatz}
   \end{subfigure}
          \begin{subfigure}[b]{0.33\textwidth}
    \includegraphics[width=\textwidth]{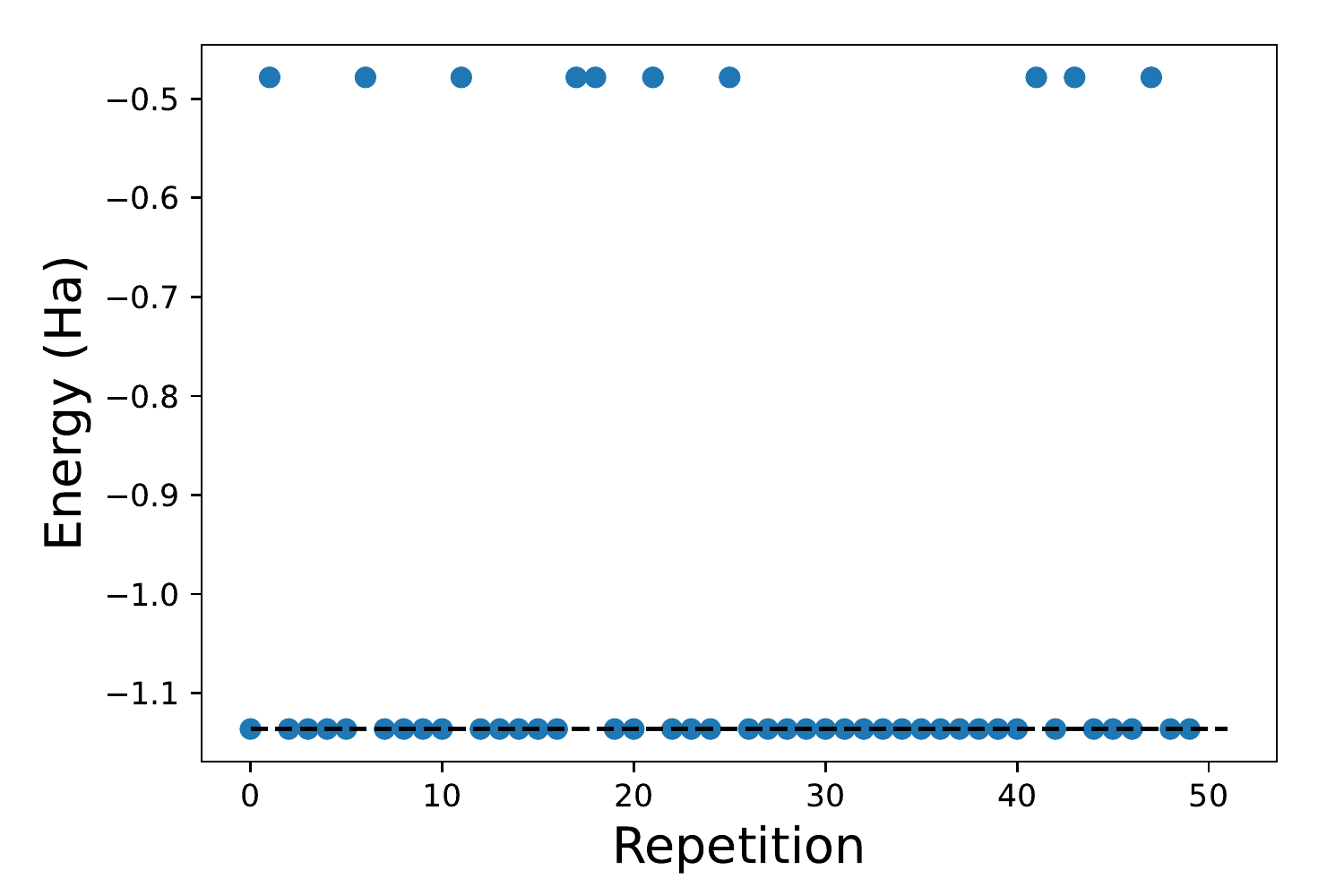}
    \caption{NFT with $R_{Y}$ ansatz}
   \end{subfigure}
          \begin{subfigure}[b]{0.33\textwidth}
    \includegraphics[width=\textwidth]{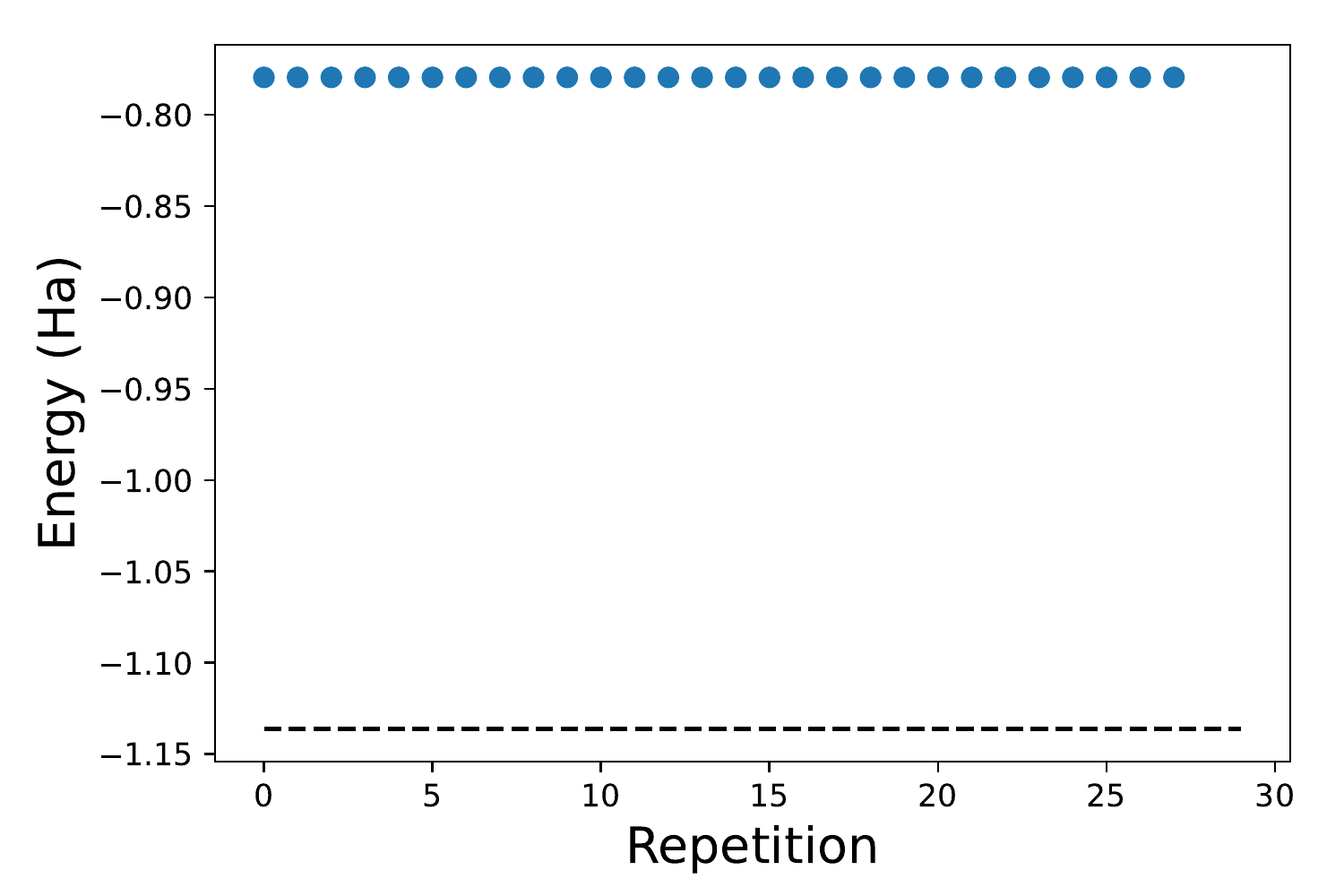} 
    \caption{Bayesian with $R_{Y}$ ansatz}
 \end{subfigure}
        \begin{subfigure}[b]{0.33\textwidth}
    \includegraphics[width=\textwidth]{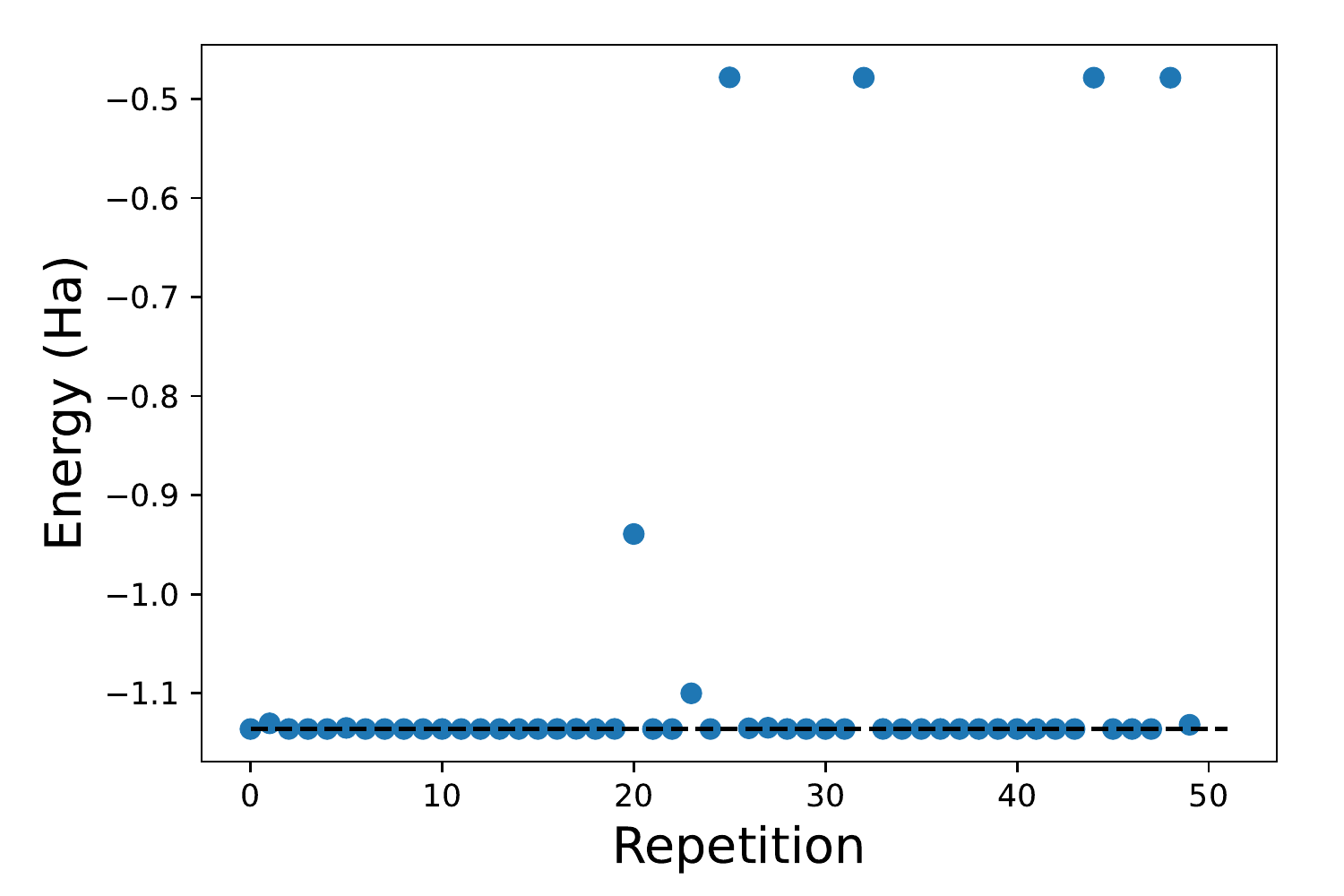}
    \caption{Modified SPSA with UCCSD ansatz}
   \end{subfigure}
          \begin{subfigure}[b]{0.33\textwidth}
    \includegraphics[width=\textwidth]{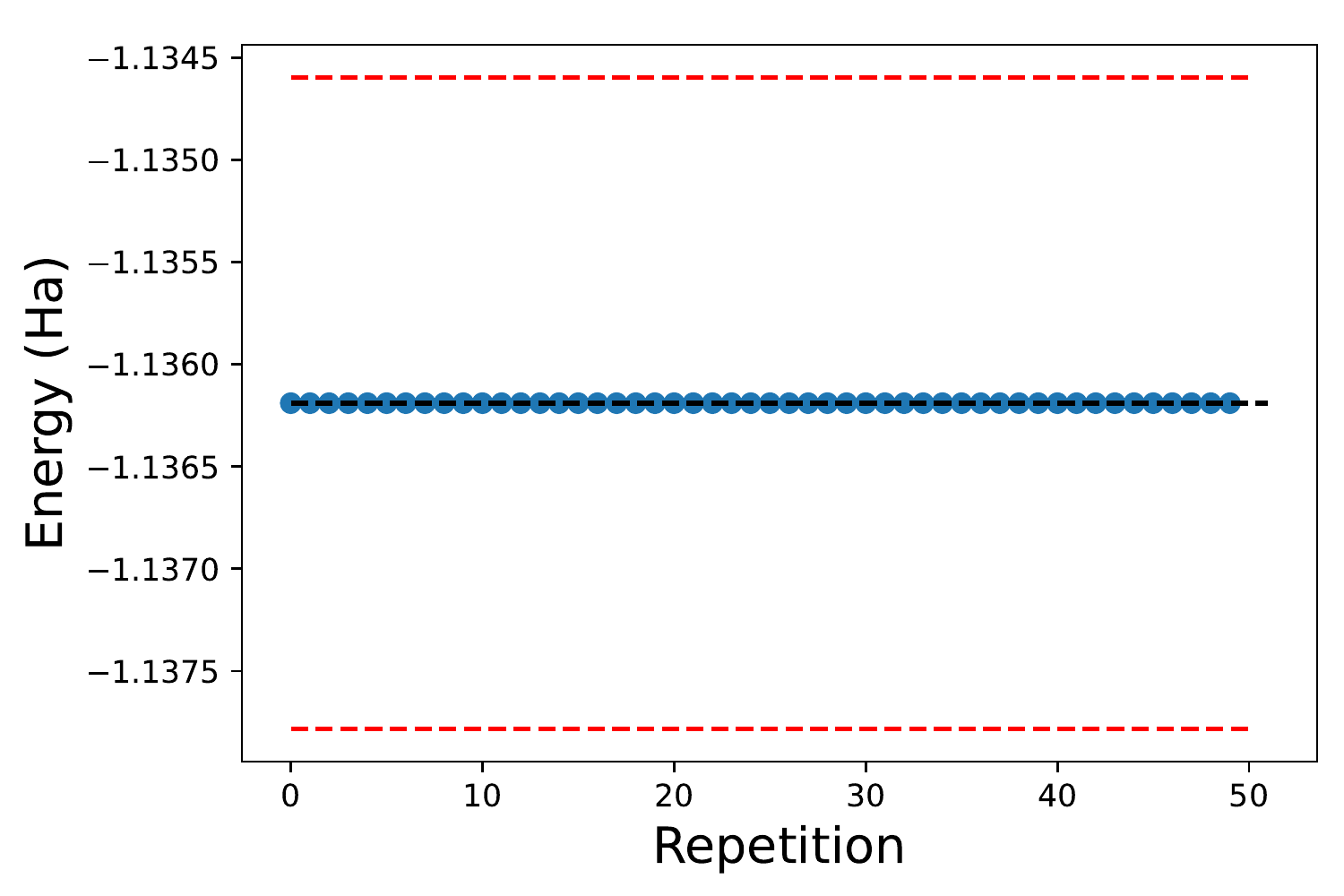}
    \caption{NFT with UCCSD ansatz}
   \end{subfigure}
          \begin{subfigure}[b]{0.33\textwidth}
    \includegraphics[width=\textwidth]{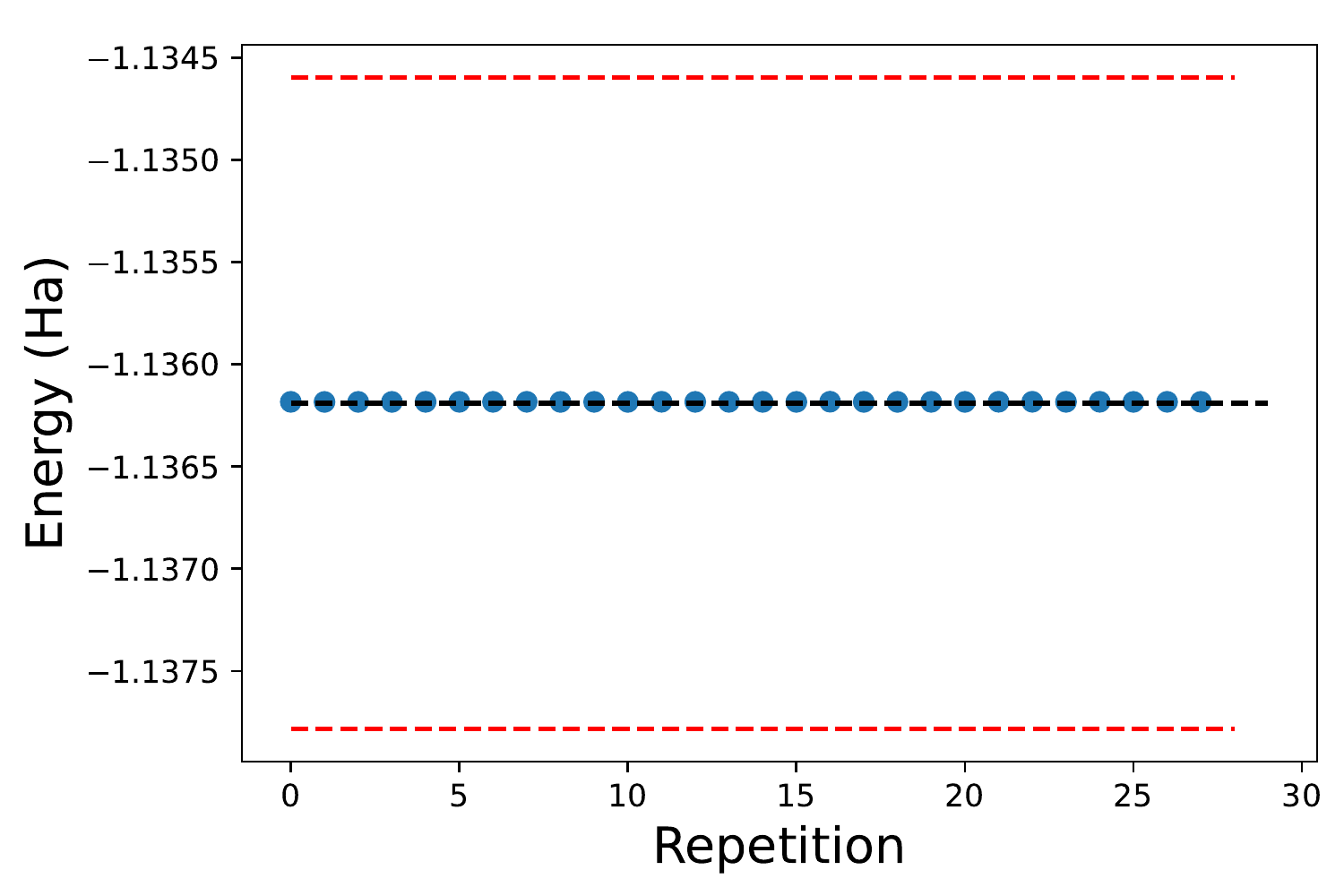}
    \caption{Bayesian with UCCSD ansatz}
     \end{subfigure}
    \caption{Results of VQE statevector simulations for several repetitions with different sets of initial parameters. The results are shown for different ansatzes and optimizers. The black dotted line marks the exact ground state energy. In figures where all resulting energies are near the exact ground state energy the chemical accuracy is indicated as red dotted line. Note the different scales.}
    \label{fig:favOptimizersW/o}
\end{figure}

\begin{figure}[h!]
    \centering
        \begin{subfigure}[b]{0.33\textwidth}
        \includegraphics[width=\textwidth]{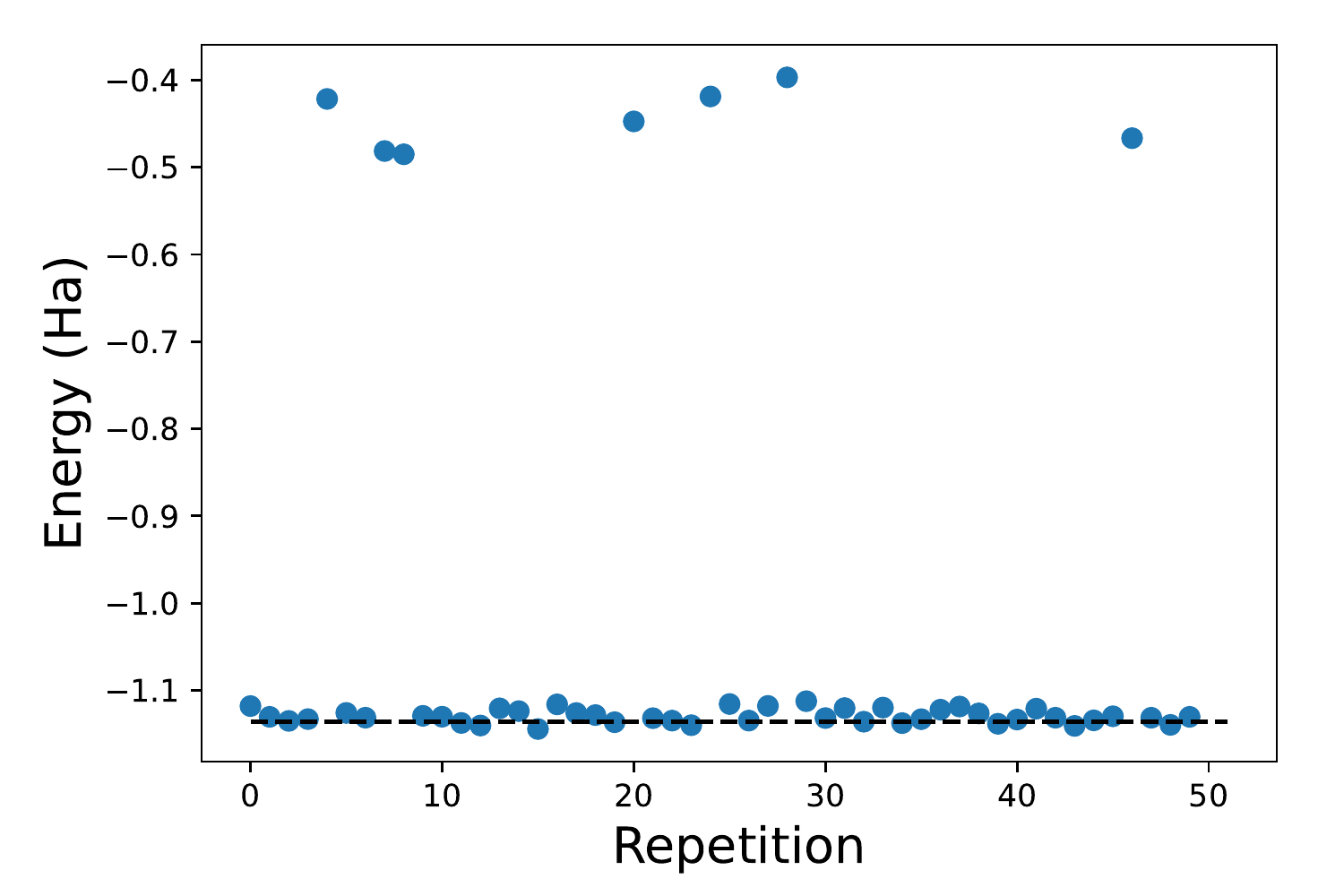}
        \caption{Modified SPSA with $R_{XYZ}$ ansatz}
        \end{subfigure}
           \begin{subfigure}[b]{0.33\textwidth}     
    \includegraphics[width=\textwidth]{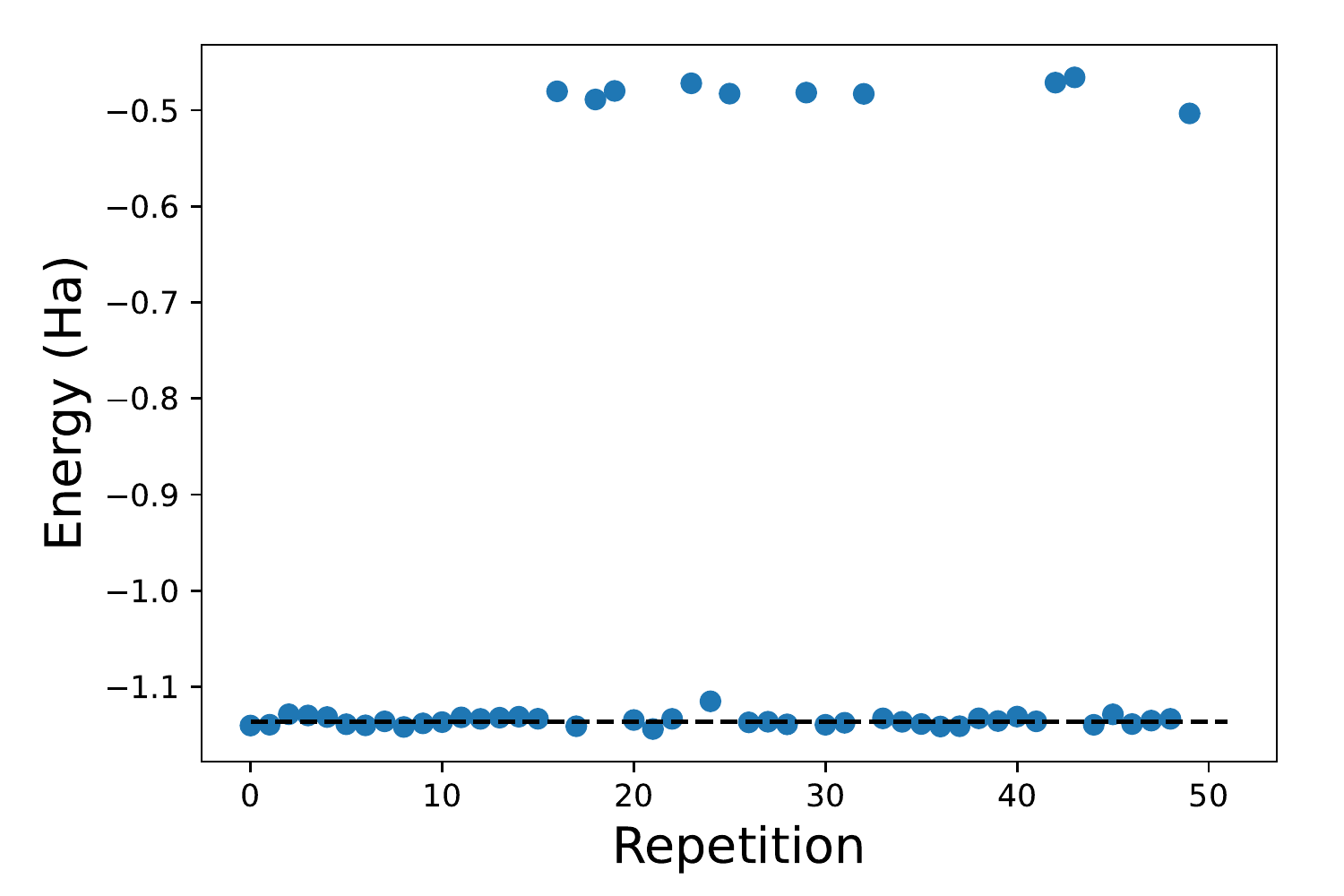} 
    \caption{NFT with $R_{XYZ}$ ansatz}
  \end{subfigure}
         \begin{subfigure}[b]{0.33\textwidth}
    \includegraphics[width=\textwidth]{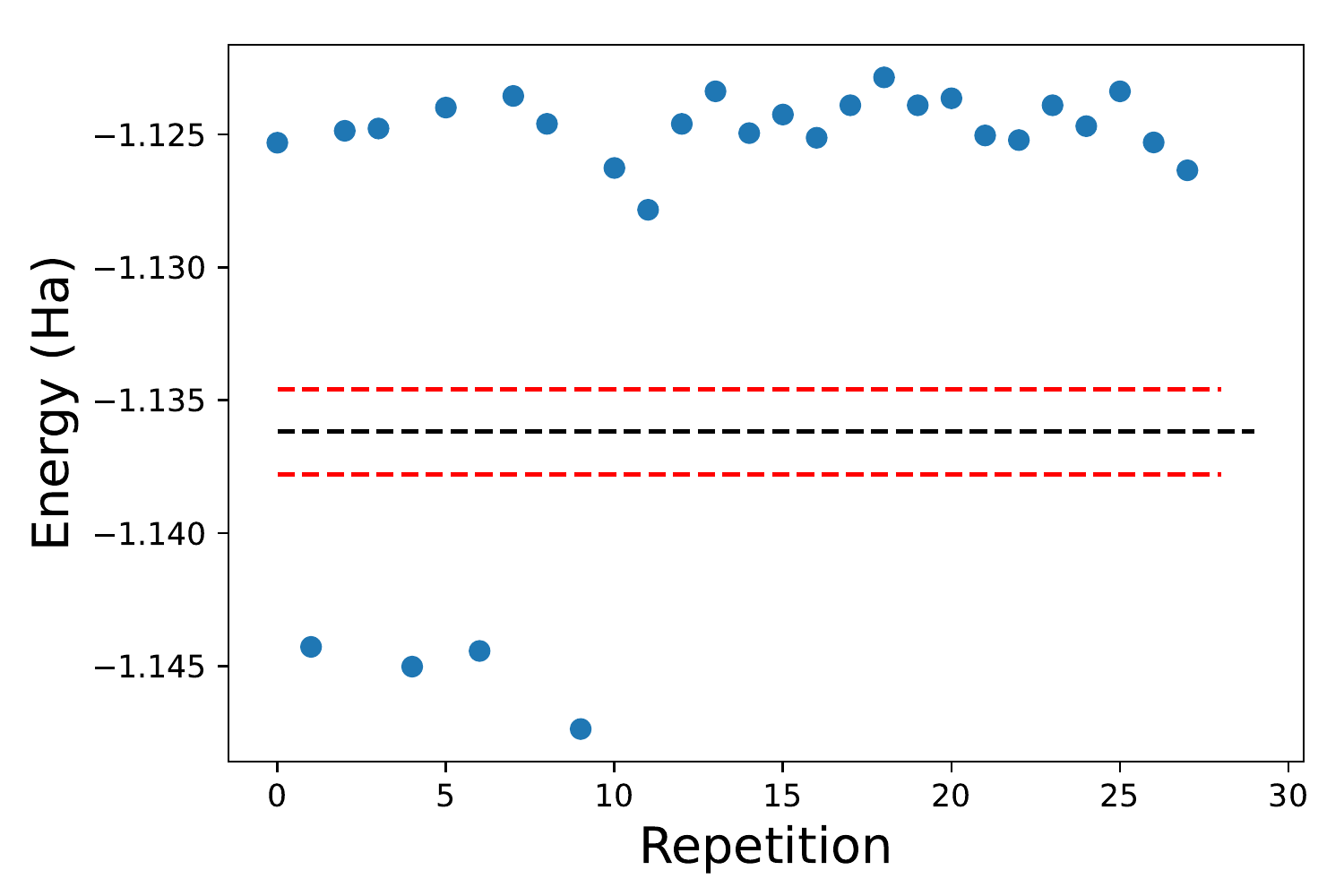}  \caption{Bayesian with $R_{XYZ}$ ansatz}
  \end{subfigure}
         \begin{subfigure}[b]{0.33\textwidth}
    \includegraphics[width=\textwidth]{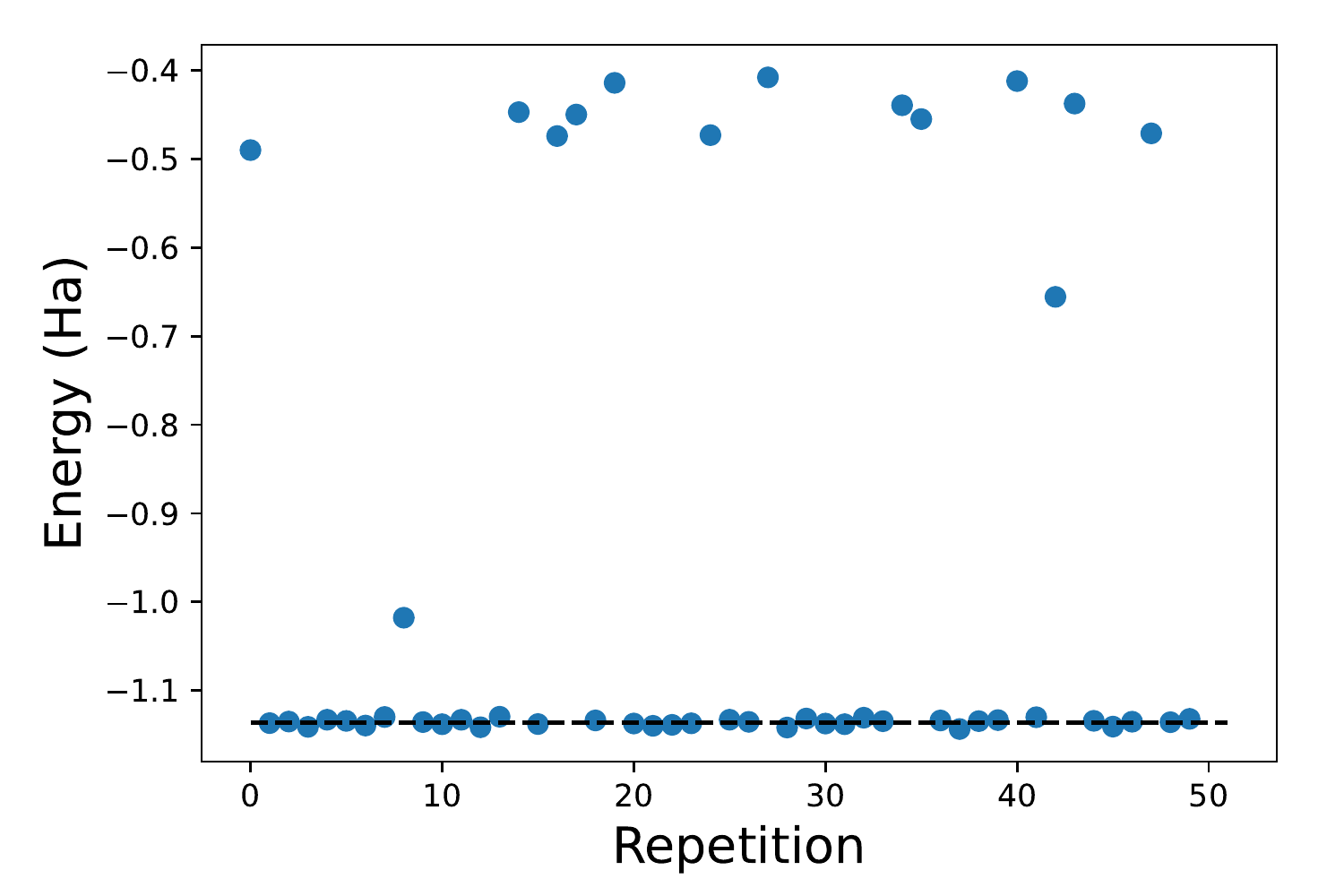}
    \caption{Modified SPSA with $R_{Y}$ ansatz}
    \end{subfigure}
           \begin{subfigure}[b]{0.33\textwidth}
    \includegraphics[width=\textwidth]{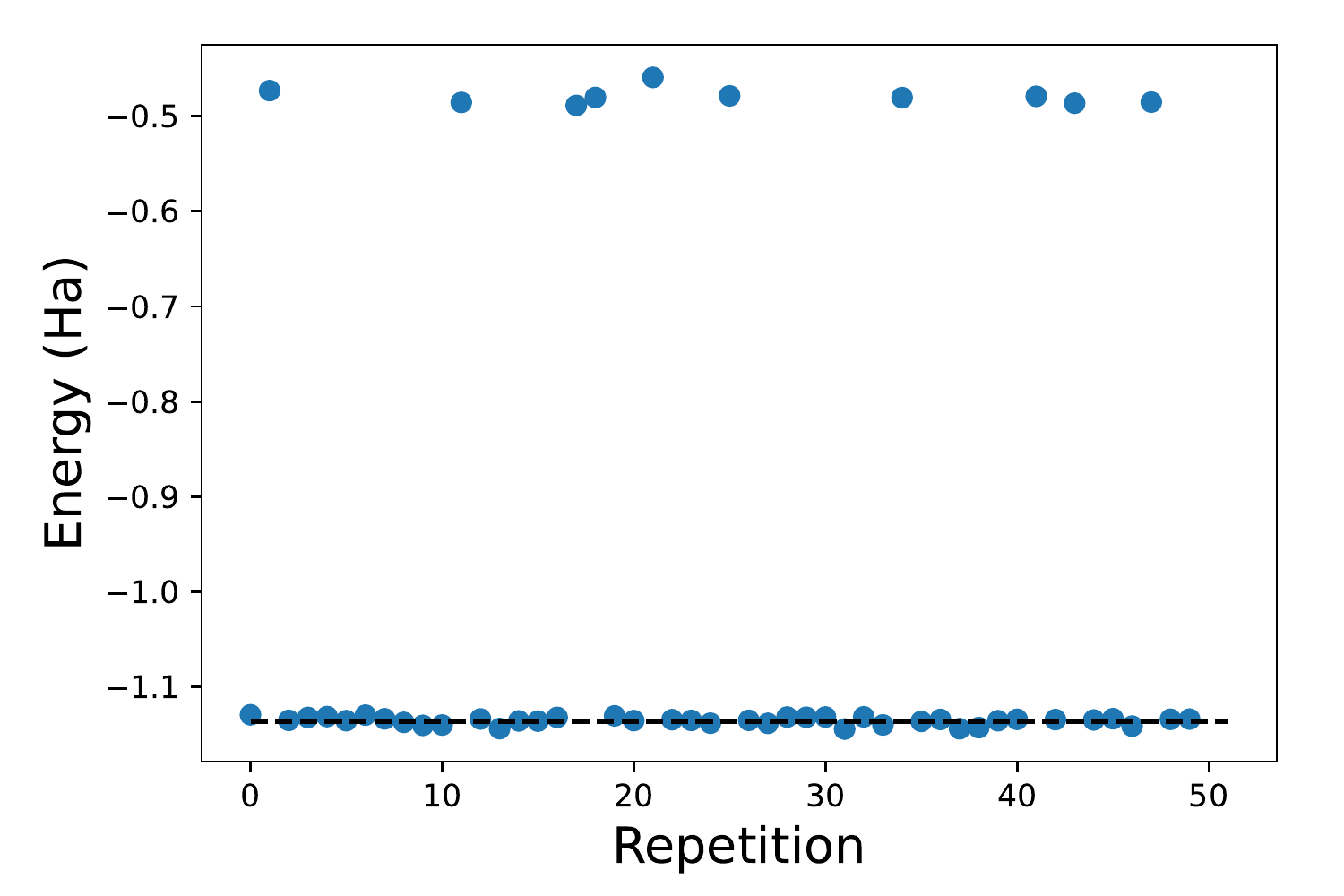}
    \caption{NFT with $R_{Y}$ ansatz}
   \end{subfigure}
          \begin{subfigure}[b]{0.33\textwidth}
  \includegraphics[width=\textwidth]{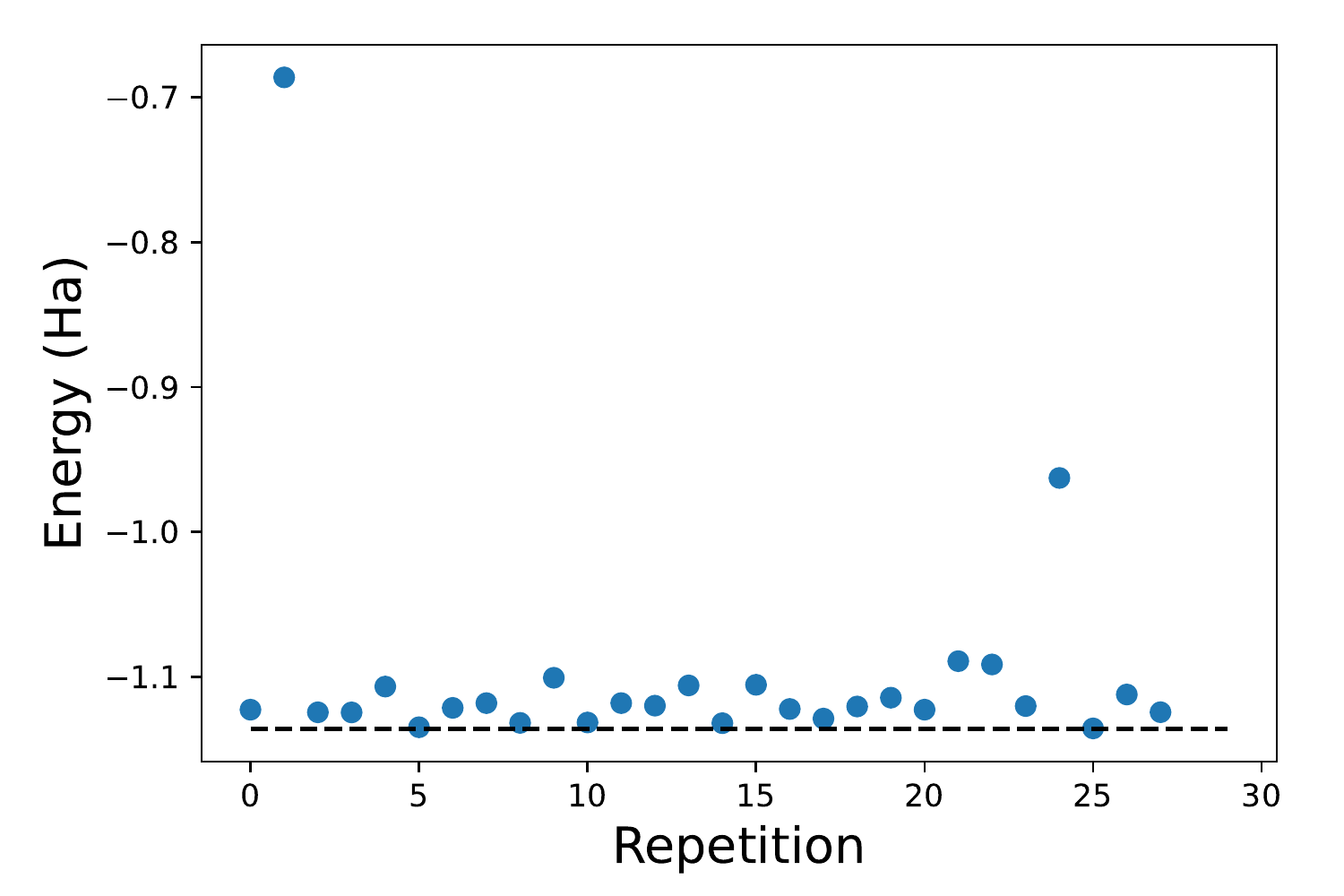} 
  \caption{Bayesian with $R_{Y}$ ansatz}
    \end{subfigure}
           \begin{subfigure}[b]{0.33\textwidth}
    \includegraphics[width=\textwidth]{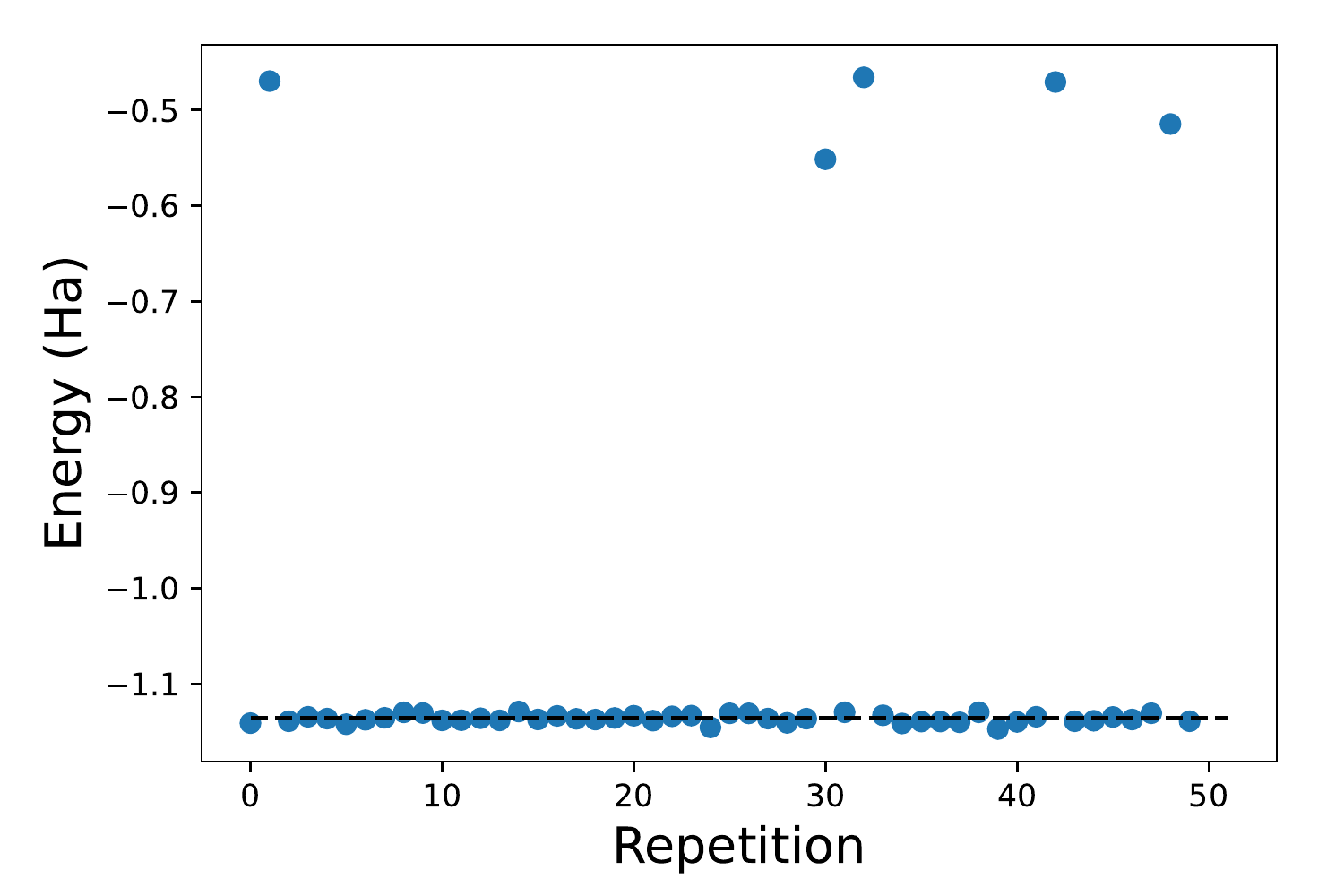}
    \caption{Modified SPSA with UCCSD ansatz}
    \end{subfigure}
           \begin{subfigure}[b]{0.33\textwidth}
    \includegraphics[width=\textwidth]{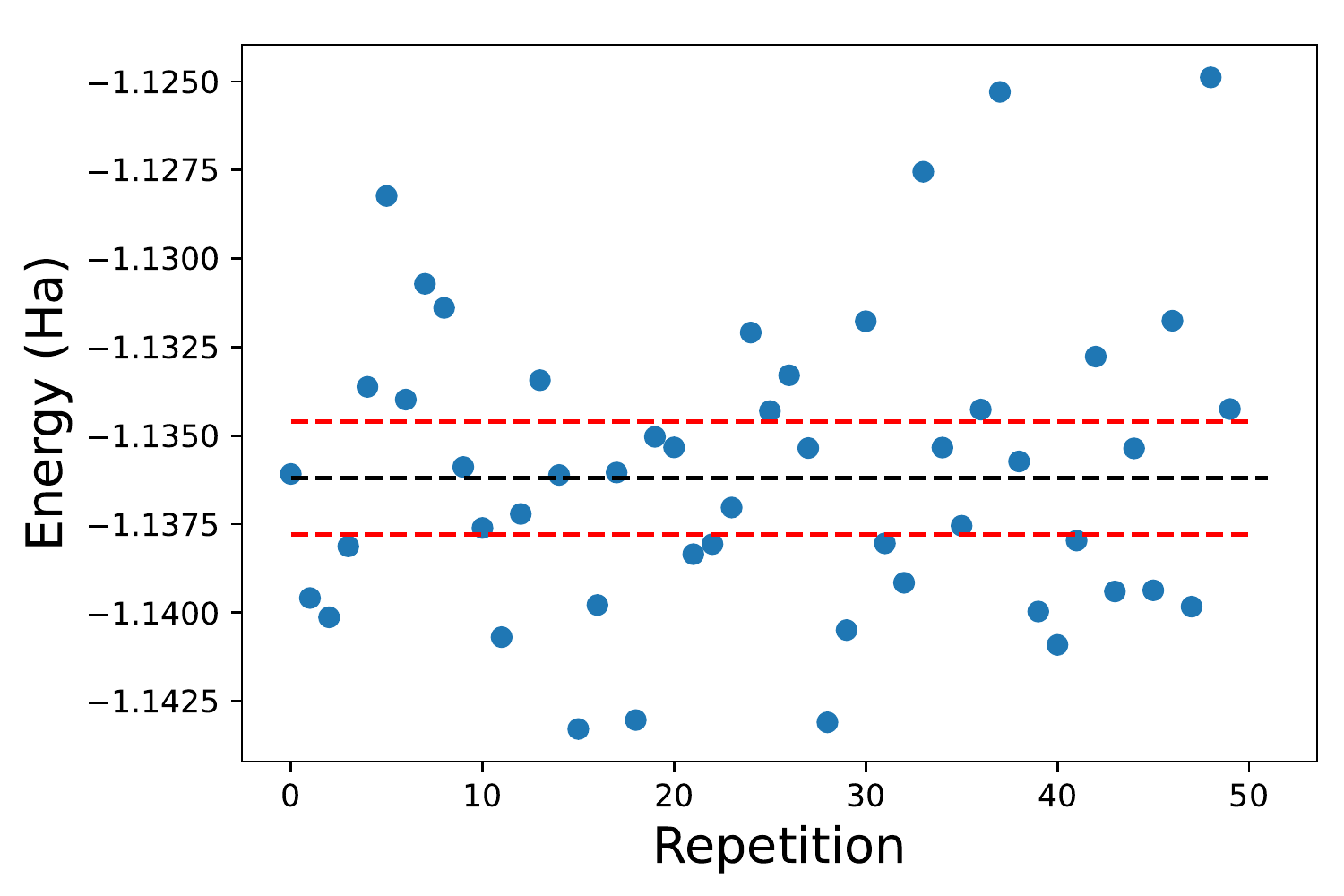}    
    \caption{NFT with UCCSD ansatz}
   \end{subfigure}
          \begin{subfigure}[b]{0.33\textwidth}
    \includegraphics[width=\textwidth]{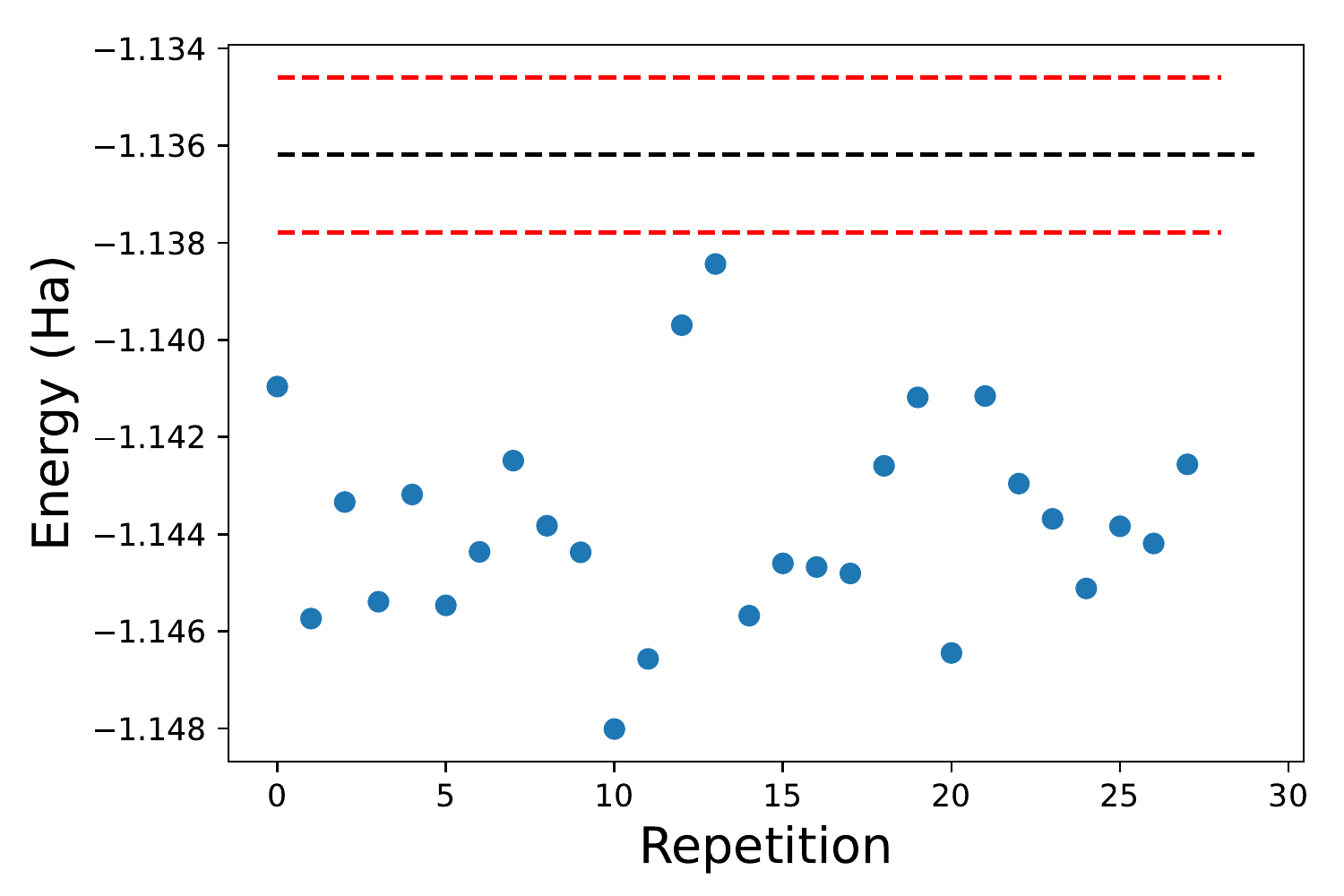}
    \caption{Bayesian with UCCSD ansatz}
     \end{subfigure}
    \caption{Results of VQE simulations including shot noise for several repetitions with different sets of initial parameters. The results are shown for different ansatzes and optimizers. The black dotted line marks the exact ground state energy. In figures where all resulting energies are near the exact ground state energy the chemical accuracy is indicated as red dotted line. Note the different scales.}  
    \label{fig:favOptimizersWn}    
\end{figure}

Three of the optimizers find good results for all tested sets of initial parameters: The modified SPSA algorithm, NFT and the Bayesian optimizer.
These are further investigated by repeating the optimization for 50 (modified SPSA and NFT), respectively 30 (Bayesian optimizer) sets of random initial parameters. In these experiments all three ansatzes are considered, i.e. $R_Y$, $R_{XYZ}$ and UCCSD, and the results are compared for statevector simulations as well as for simulations including shot noise. For the latter 1024 shots are used.

The obtained minimal energies for the statevector simulations are shown in figure~\ref{fig:favOptimizersW/o}. The results of the Bayesian optimizer are independent of the initial parameters, while for NFT and the modified SPSA two energy levels appear. This discrete behavior has also been observed in other works, e.g. in \cite{mihalikova_best-practice_2022}, and may indicate excited energy levels of the Hamiltonian.
Compared to NFT, the modified SPSA shows more fluctuations in its results around these two levels due to its stochastic nature. 
For the modified SPSA the number of results far away from the exact ground state energy is smaller than for NFT, but the accuracy of the results that are near the exact ground state energy is higher with NFT. Especially the Bayesian optimizer shows a strong dependence on the ansatz: For the $R_{XYZ}$ and UCCSD ansatzes it gives precise and stable results, which lie within the chemical accuracy of $1.6 \cdot 10^{-3}$~Ha~\cite{mcardle_quantum_2020}. However, for the $R_Y$ ansatz it clearly gives insufficient results.

The results with shot noise are shown in figure~\ref{fig:favOptimizersWn}. 
For all three optimizers the results spread around the original values, also below the true ground state energy. This is caused by the fact that the Hamiltonian in \ref{H2Ham} is a sum of different Pauli strings, which are measured individually. Since the measurement can yield a result other than its analytical expectation value, the sum of these terms may be smaller than the exact ground state energy. 
In presence of noise, the Bayesian optimizer gives results close to the exact ground state energy for all three ansatzes. Compared to the modified SPSA and NFT, the results for the Bayesian optimizer yield the lowest number of points far away from the exact ground state energy. However, for the UCCSD ansatz the results are distributed below the ground state energy. For this ansatz the NFT optimizer provides the best results. Furthermore, the optimization process is significantly faster with NFT and it also needs much less iterations to converge. Due to that, and because it shows a good performance for all ansatze, we choose the NFT optimizer for the subsequent noise studies. 

\subsubsection{Impact of noise}

\graphicspath{{./imagesNoise/pdf}}

In this section the effects of the different kinds of noise are investigated. For this, we run the VQE algorithm for all three ansatzes and with several noise intensities, which we can vary in simulations. We use the NFT optimizer and for each setup 30 repetitions with one set of initial parameters are performed. The noise intensities, corresponding to the probabilities of noise described in section \ref{noise theory}, are chosen to cover the typical values of current IBM Quantum Falcon r5.11 version 1 superconducting devices by at least one order of magnitude.

\begin{figure}[h!]
    \centering
    \begin{subfigure}[b]{0.49\textwidth}
    \includegraphics[width=\textwidth]{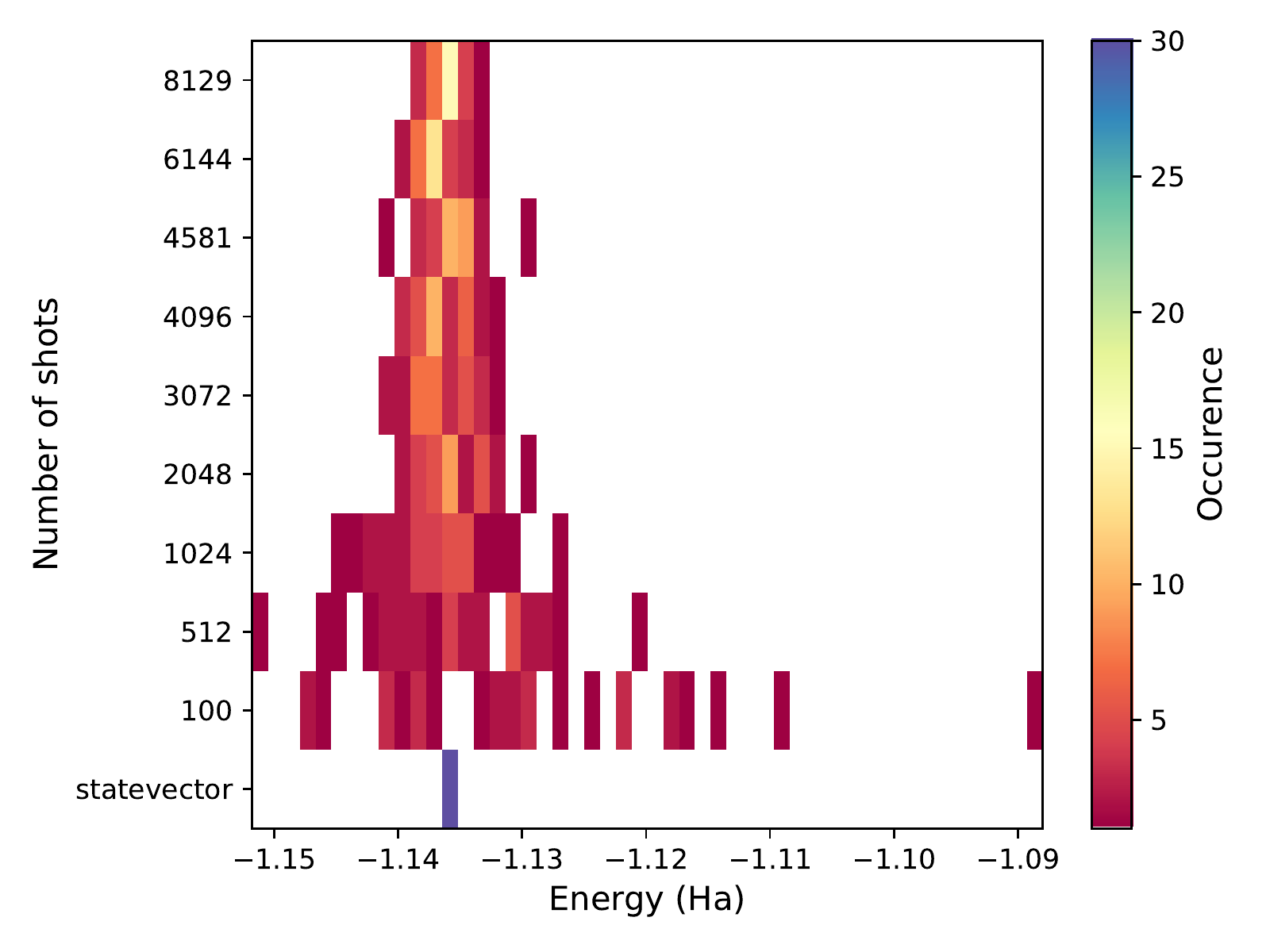}
    \caption{Shot noise with $R_{XYZ}$ ansatz}
    \label{fig:histograms_shots}
    \end{subfigure} \hfill
    \begin{subfigure}[b]{0.49\textwidth}
    \includegraphics[width=\textwidth]{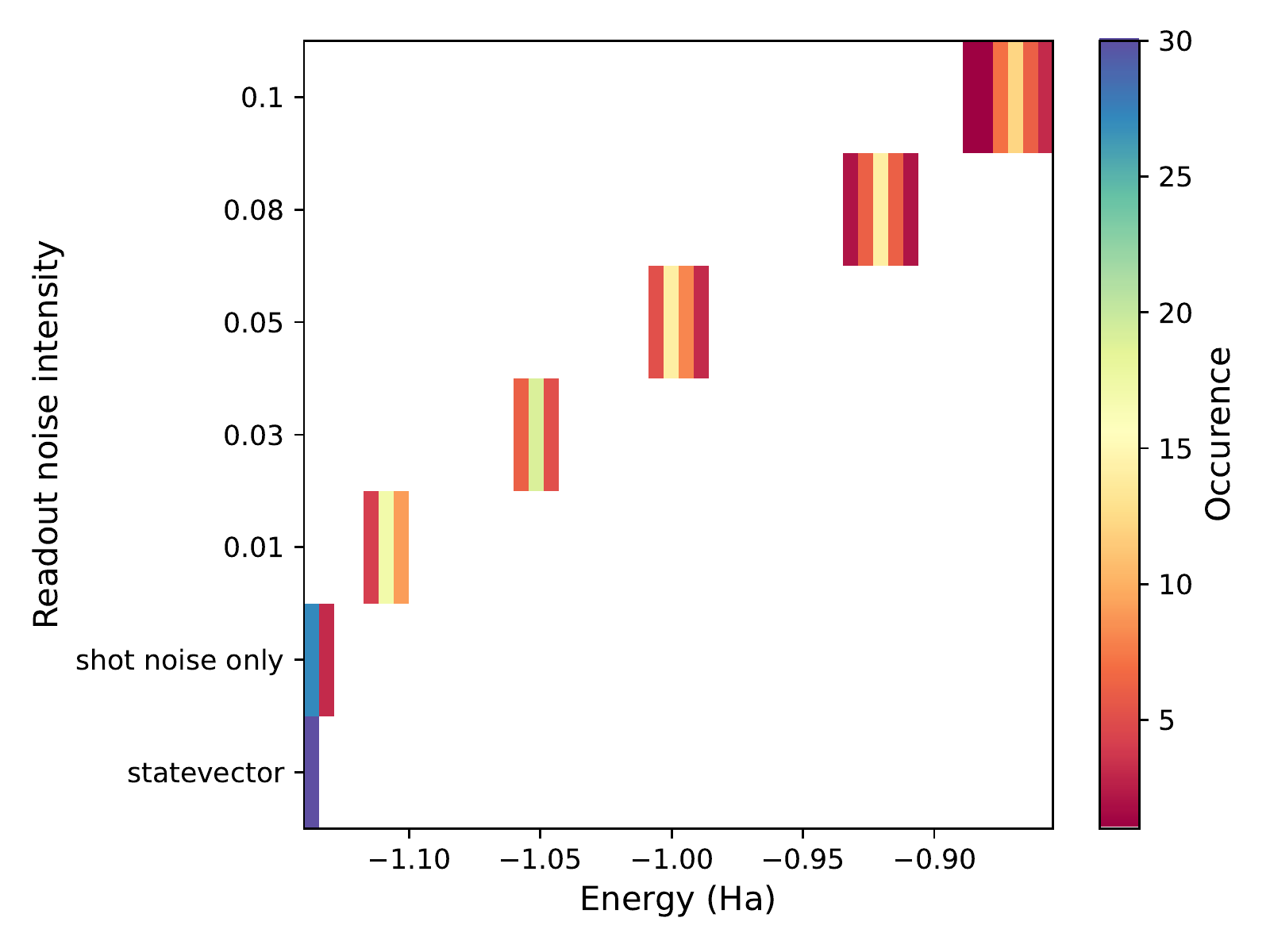}
    \caption{Readout noise with $R_{XYZ}$ ansatz, 1024 shots}
    \label{fig:histograms_readout}
    \end{subfigure}
    \begin{subfigure}[b]{0.49\textwidth}
    \includegraphics[width=\textwidth]{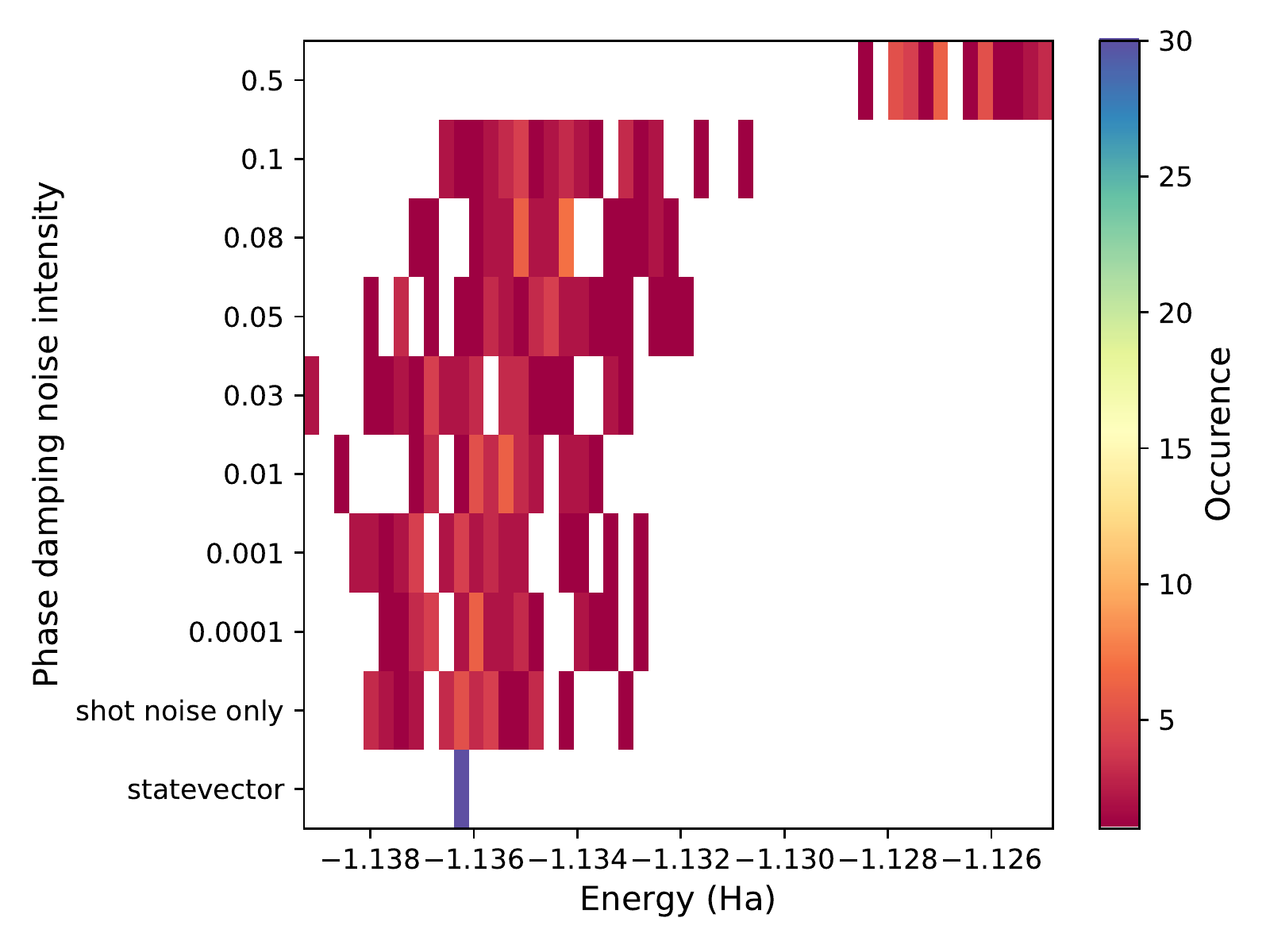} 
    \caption{Phase damping noise with $R_{XYZ}$ ansatz, 8192 shots}    
    \label{fig:histograms_phase_rot}
    \end{subfigure} \hfill
    \begin{subfigure}[b]{0.49\textwidth}
    \includegraphics[width=\textwidth]{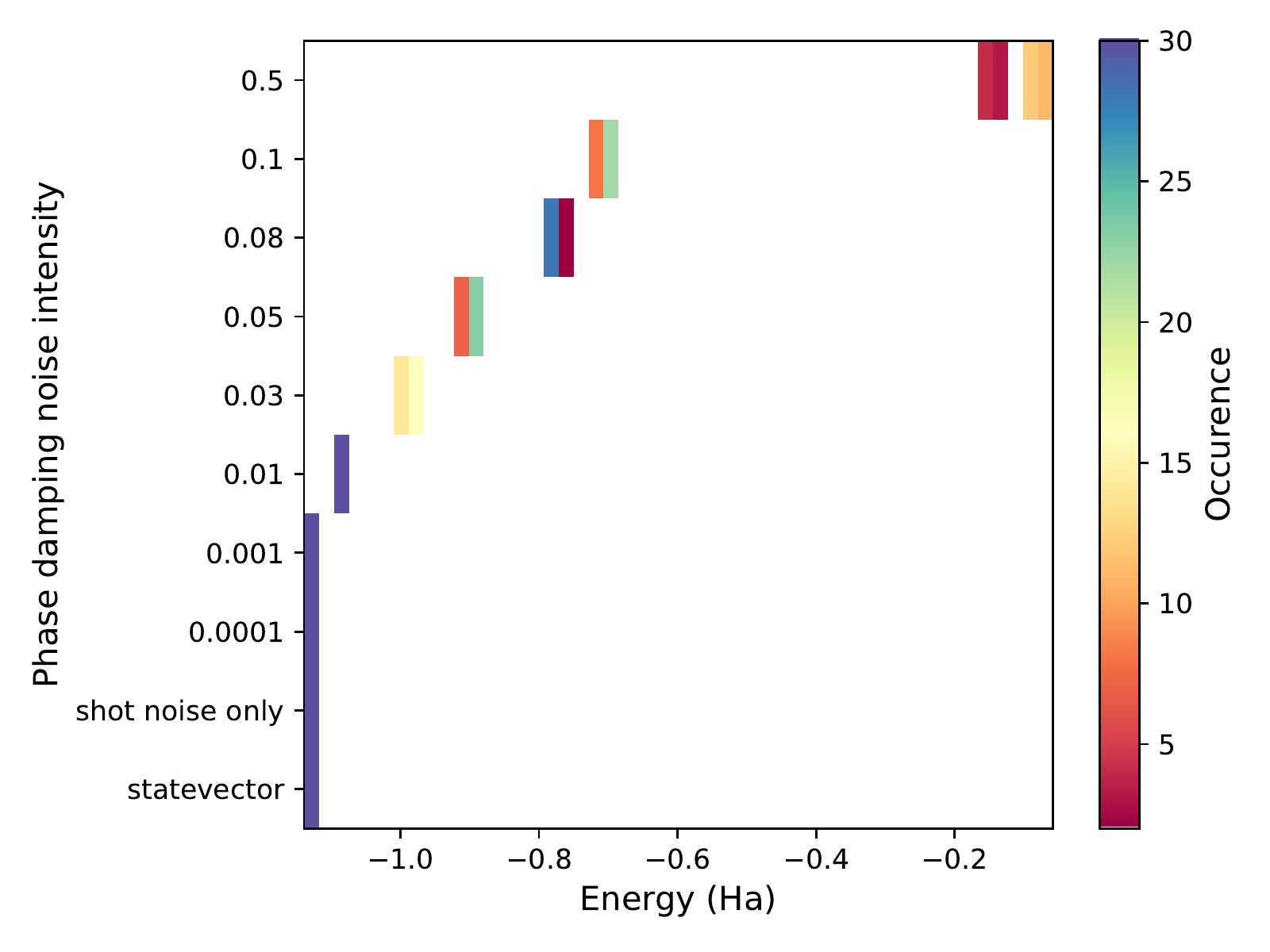}
    \caption{Phase damping noise with UCCSD ansatz, 8192 shots}     
    \label{fig:histograms_phase_ucc}
    \end{subfigure}
    \caption{Histograms of the resulting energies of 30 repetitions of VQE for simulations with various noise intensities of different kinds of noise. Zero entries are depicted as white. Note the different scales. }
    \label{fig:histograms}
\end{figure}

Four exemplary heat maps from our noise studies are depicted in Figure~\ref{fig:histograms}. They show histograms of the obtained energy values for different noise intensities or shot numbers. For reference, also the noise-free energy as calculated from statevector simulations is added. The exact ground state energy lies within the energy bin from the statevector simulation for all cases. Figure~\ref{fig:histograms_shots} shows the influence of shot noise on the accuracy of the results for the $R_{XYZ}$ ansatz. A decreasing number of shots leads to a broadening of the distribution and energy values below the exact ground state energy, as already seen in section \ref{Optimizers}. Also for other ansatzes and with other noise sources included, only the fluctuations around the mean energy value change with the shot number, but the mean value itself remains the same within the uncertainties.

\begin{figure}[h!]
    \centering
       \begin{subfigure}[b]{0.49\textwidth}    
    \includegraphics[width=\textwidth]{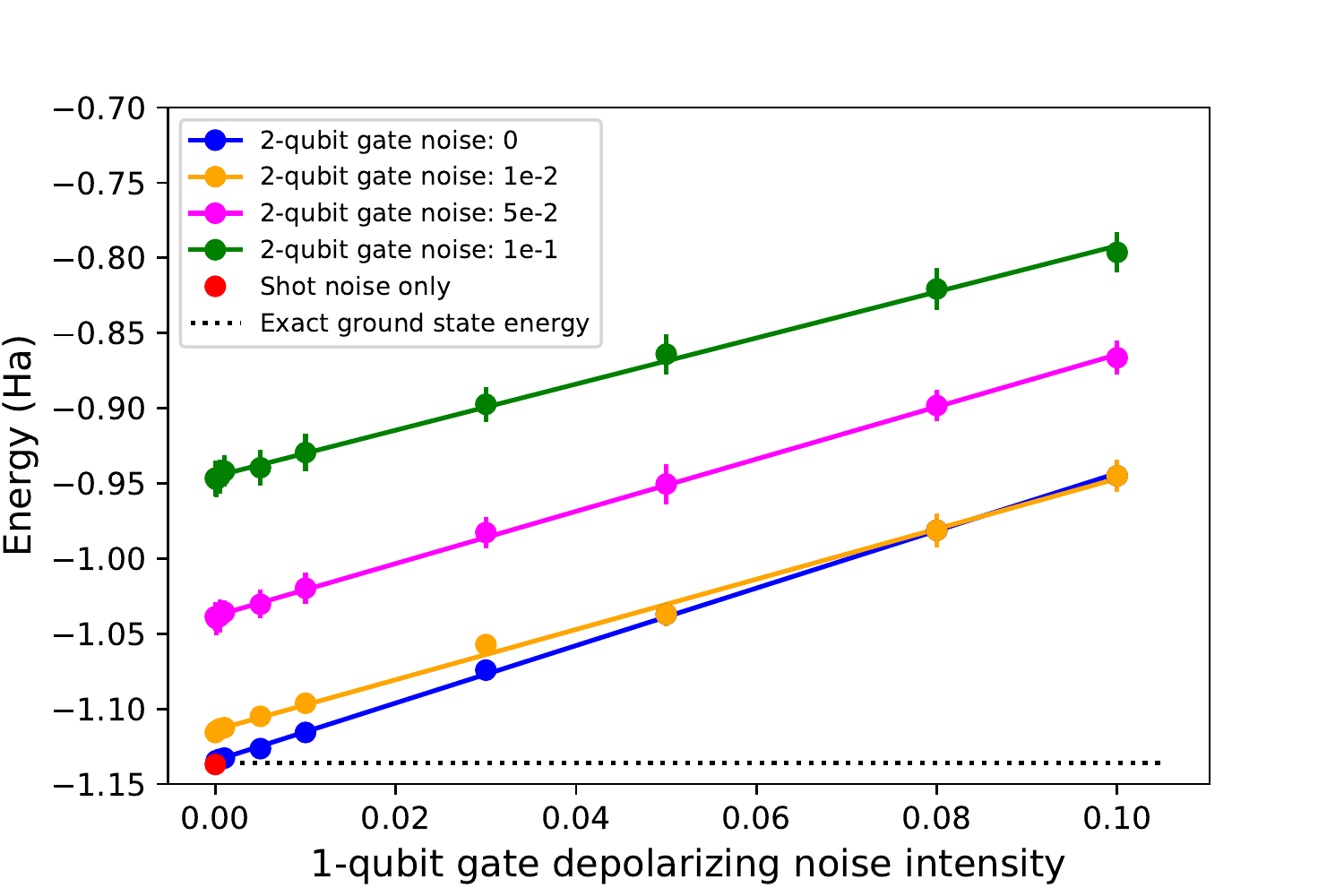} \end{subfigure}
       \begin{subfigure}[b]{0.49\textwidth}     
    \includegraphics[width=\textwidth]{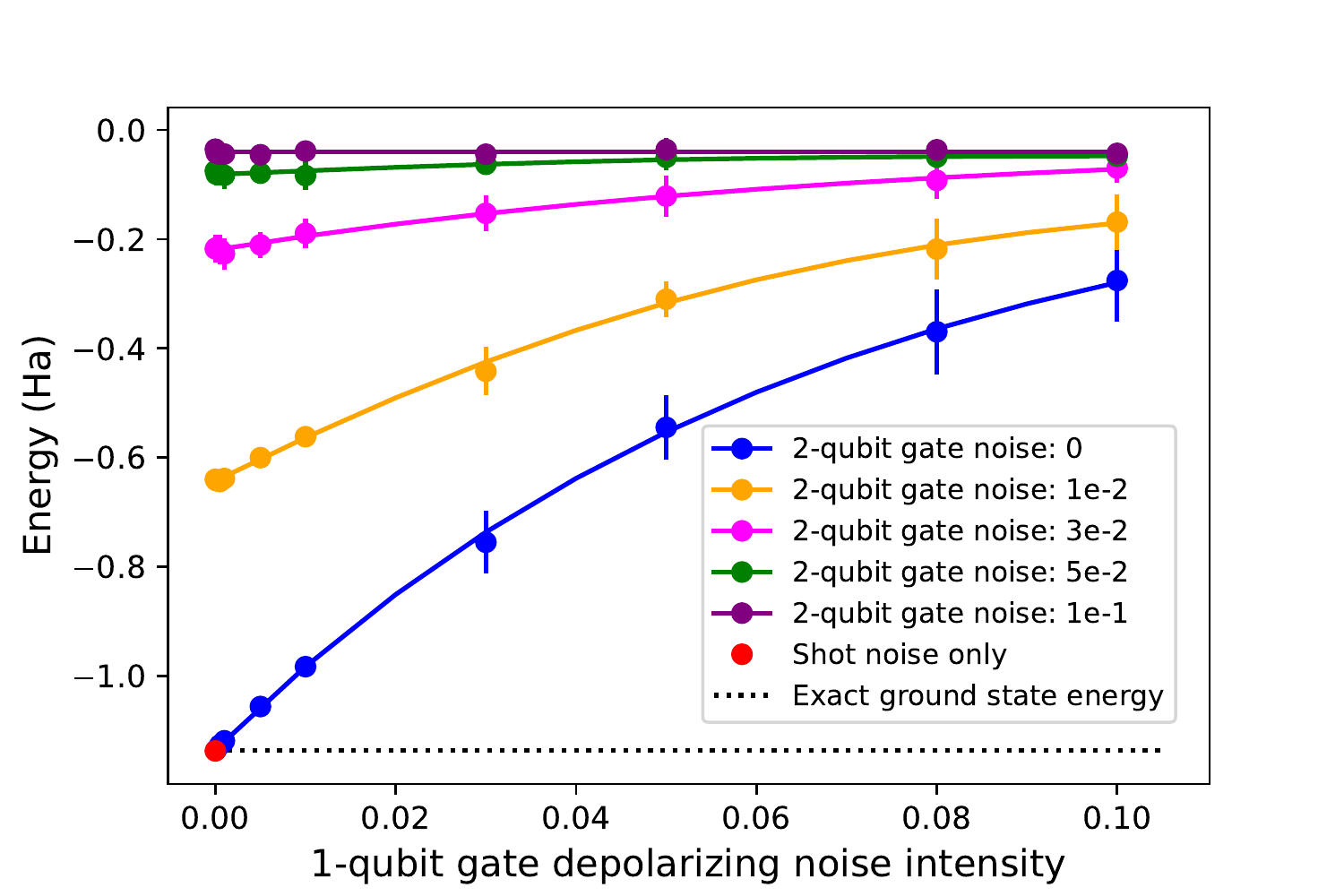}    
    \end{subfigure}
    \begin{subfigure}[b]{0.49\textwidth} 
    \includegraphics[width=\textwidth]{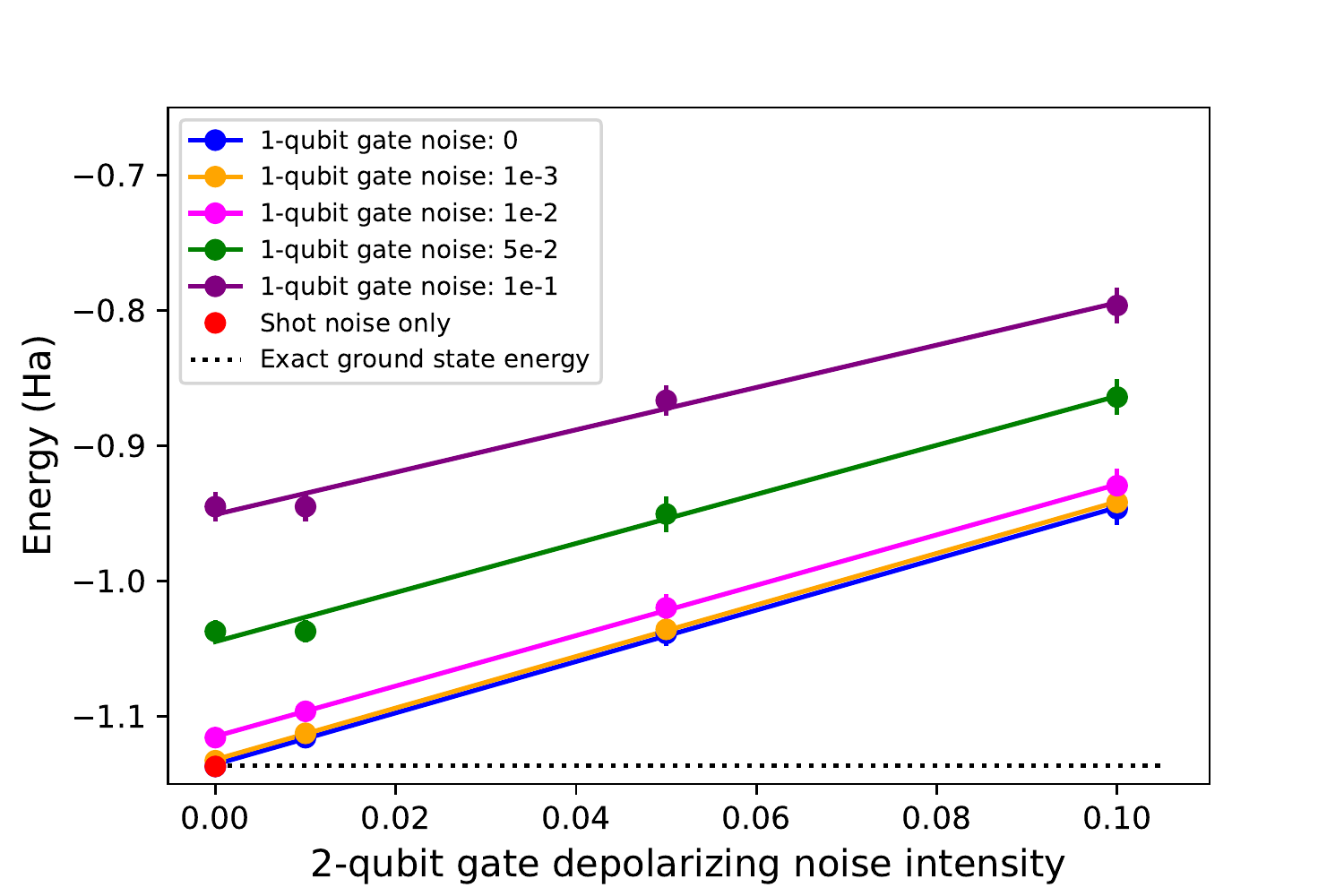} 
    \end{subfigure}
    \begin{subfigure}[b]{0.49\textwidth} 
     \includegraphics[width=\textwidth]{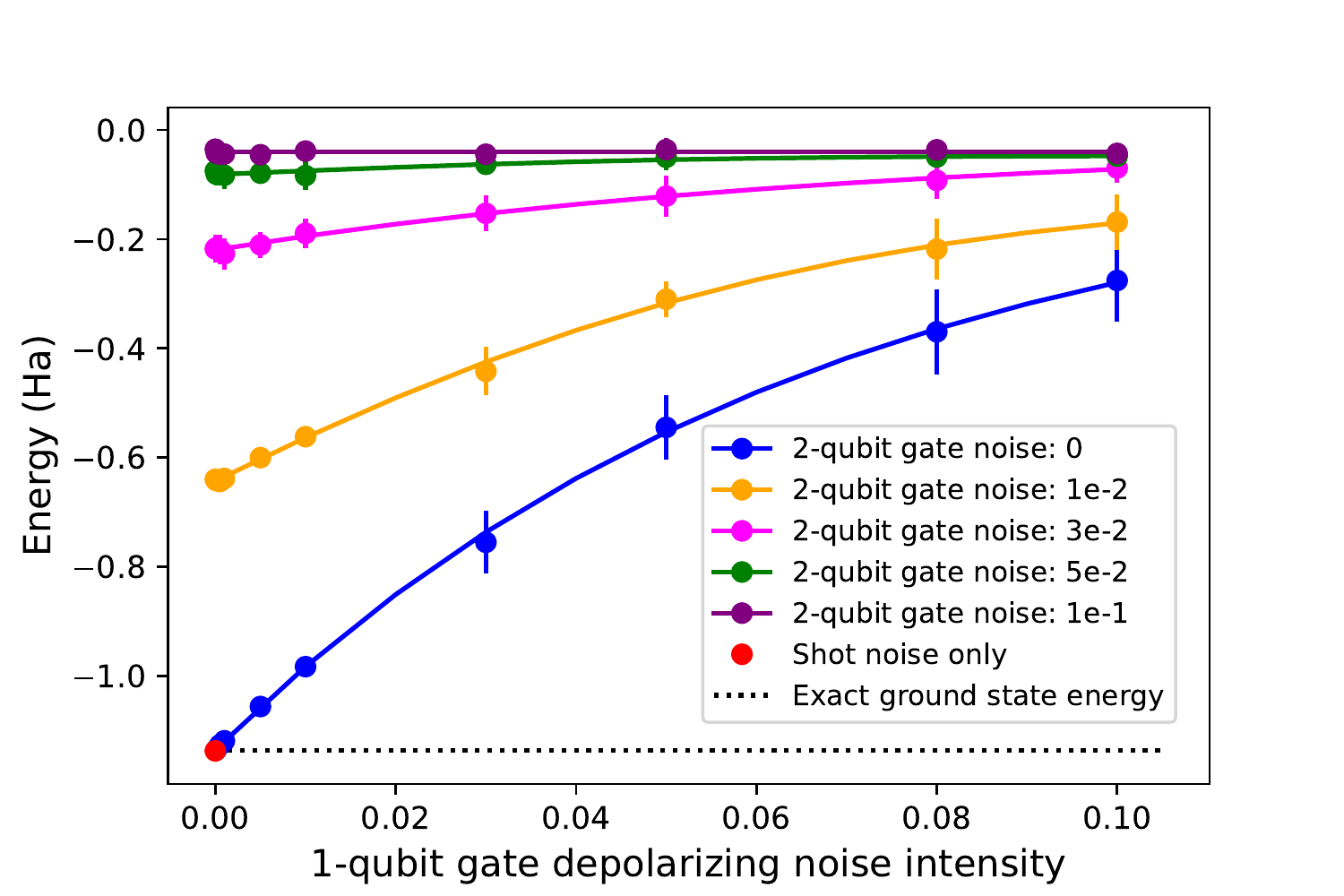}   
     \end{subfigure}
     \begin{subfigure}[b]{0.49\textwidth} 
    \includegraphics[width=\textwidth]{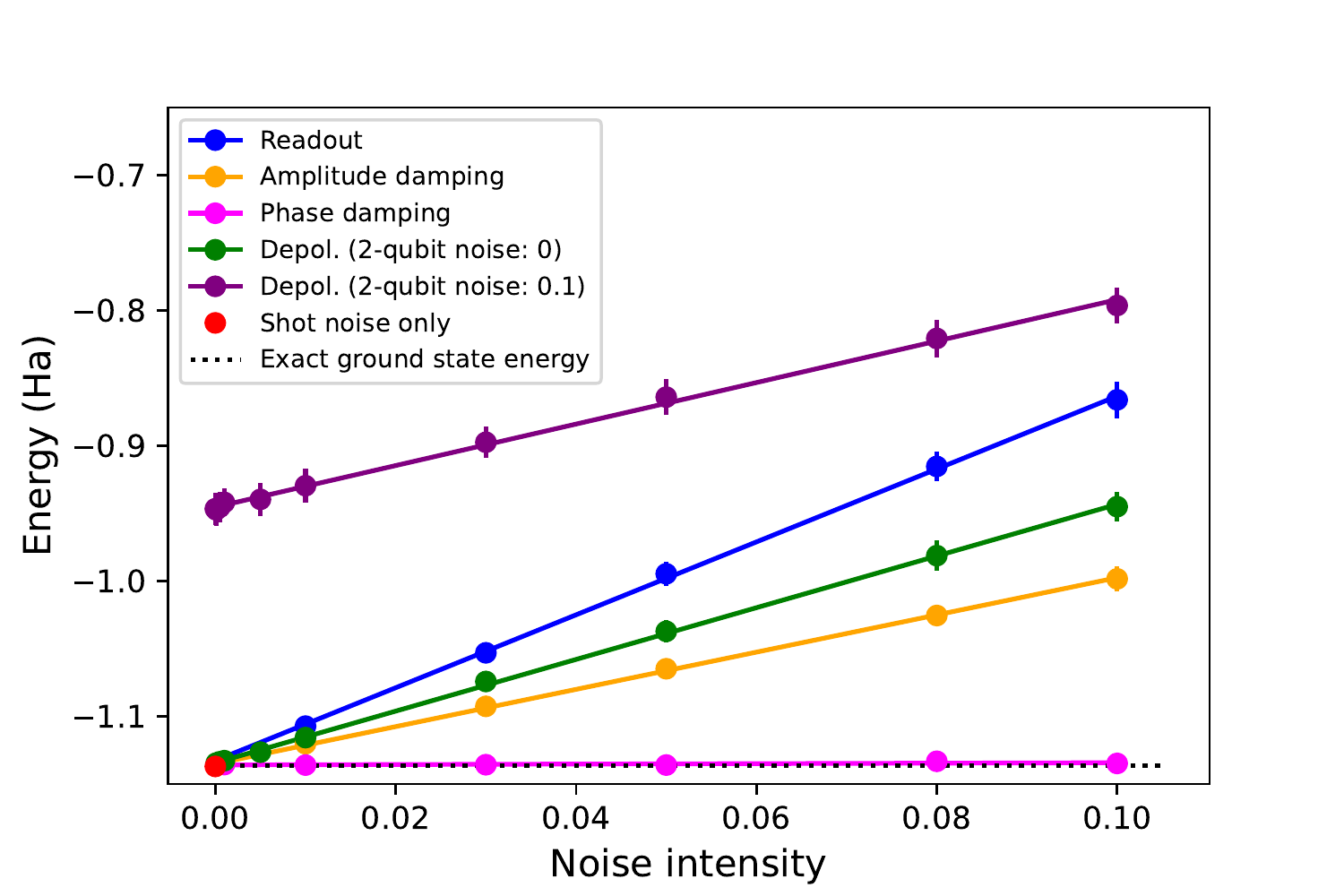} 
    \end{subfigure}
    \begin{subfigure}[b]{0.49\textwidth} 
     \includegraphics[width=\textwidth]{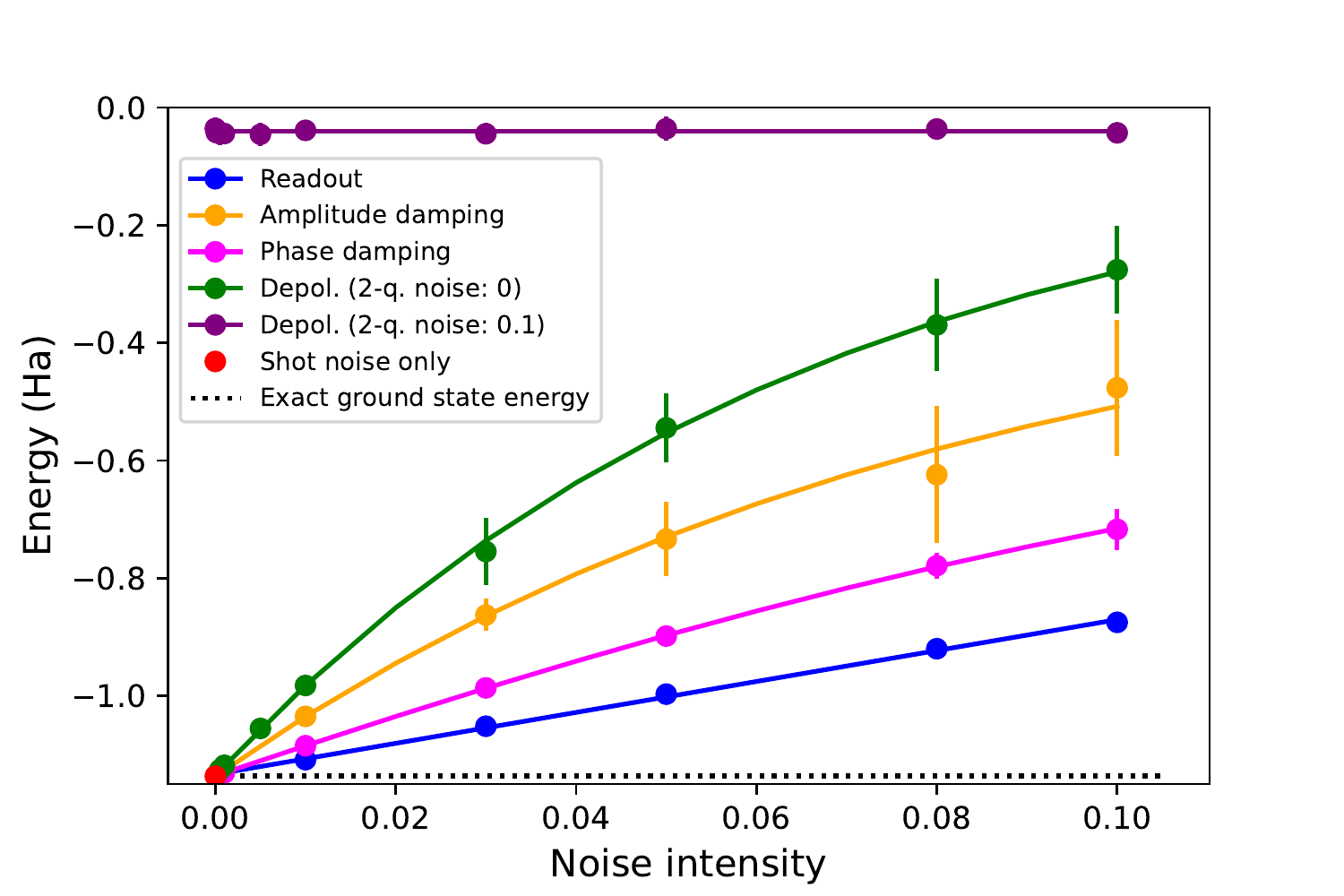} 
     \end{subfigure}
    \caption{Dependence of the energy shift from the noise intensity for simulations with the $R_{XYZ}$ ansatz (left) and the UCCSD ansatz (right). The top plots show the mean energies as function of the noise intensity of 1-qubit gate depolarizing noise for different values of 2-qubit gate depolarizing noise. The middle row shows the mean energies as function of the noise intensity of 2-qubit gate depolarizing noise for different values of 1-qubit gate depolarizing noise. The bottom plots show noise-energy curves for different kinds of noise. The error bars indicate the standard deviation of the 30 simulation repetitions. A linear function is used to fit the energy values obtained with the $R_{XYZ}$ ansatz as well as to fit the readout noise curve for the UCCSD ansatz. For the other noise sources, a Gauss error function is used.}
    \label{fig:depHwe}
\end{figure}

Additional sources of noise introduce a shift of the energies towards higher values, as can be seen in figure~\ref{fig:histograms_readout} for the $R_{XYZ}$ ansatz with readout noise. 
The extent of this energy shift highly depends on the type of noise. This becomes apparent when comparing figure~\ref{fig:histograms_readout} and figure~\ref{fig:histograms_phase_rot}, which show the obtained energies for different strengths of readout and phase damping noise, respectively. The $R_{XYZ}$ ansatz is used in both cases.
For other ansatzes, the impact of the different noise types may significantly change. Figure~\ref{fig:histograms_phase_ucc} shows the impact of phase damping noise for the UCCSD ansatz, which is much larger compared to figure~\ref{fig:histograms_phase_rot}.

In a next step, we investigate these noise effects in more detail. For this, we plot the mean energy values of the 30 repetitions as a function of the noise intensity. Since shot noise leaves the mean energy value unaffected, a shot number of 1024 is sufficient for the noise simulations.
To derive a mathematical formulation for the noise-energy relationship, we fit the distributions with appropriate fit functions. Figure~\ref{fig:depHwe} shows the results for the $R_{XYZ}$ and UCCSD ansatzes. These compare the fitted curves between the four different noise types, i.e. readout, amplitude damping, phase damping and depolarizing noise. Since we model depolarizing noise for 1 and 2-qubit gates separately, the results in figure~\ref{fig:depHwe} also show comparisons of noise-energy curves with varying combinations of depolarizing noise intensities. For the considered devices the typical intensities for 1 and 2-qubit gates lie around 0.001 and 0.01, respectively~\cite{ibmq}. Additionally, the mean energy for runs with shot noise only is shown.
 
\begin{figure}[h!]
    \centering
    \includegraphics[width=0.6\textwidth]{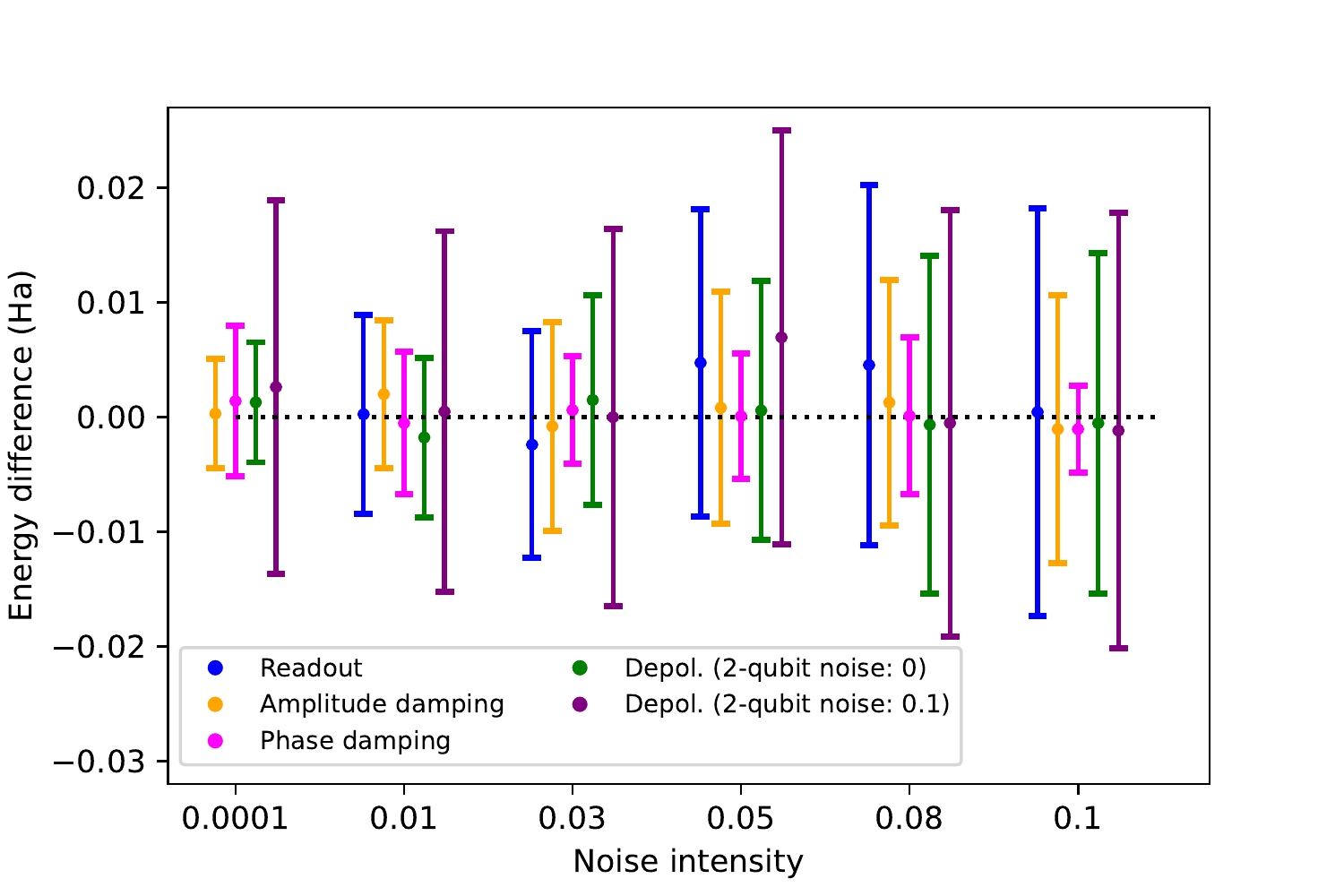}
    \caption{Difference of the mean energy values resulting from simulations with the $R_{XYZ}$ and the $R_Y$ ansatz for six noise intensities and different kinds of noise.}
    \label{fig:hweComp}
\end{figure}

For the hardware-efficient $R_{XYZ}$ ansatz we find a linear relationship between the mean values of energies and the noise intensity for all noise types with high accuracy. 
Among the different curves, readout noise has the largest gradient. Together with the fact that the error rates of readout noise of superconducting qubits are typically highest, one can expect this type of noise to have the biggest influence on the unmitigated result of VQE on real hardware for this ansatz. For a typical readout noise intensity of around 0.03~\cite{ibmq} an energy shift of about 7~\% with respect to the statevector value is observed. If the readout noise increases by a factor of about 3, i.e. to a readout noise intensity around 0.1, a deviation of approximately 24~\% is obtained. Especially noteworthy is the small influence of the phase damping noise, with typical values around 10$^{-3}$ and under 0.2\% energy shift relative to the exact ground state energy for noise intensities up to 10\%. We observe very similar results for the $R_Y$ ansatz. The differences of the results obtained with the $R_{XYZ}$ and the $R_Y$ ansatz are within the standard deviation of the energy distributions for all types of noise, as shown in figure \ref{fig:hweComp}.
This indicates that the effect of noise does not depend on the type of rotation in the hardware-efficient ansatzes.

\begin{figure}[h!]
    \centering
    \includegraphics[width=0.56\textwidth]{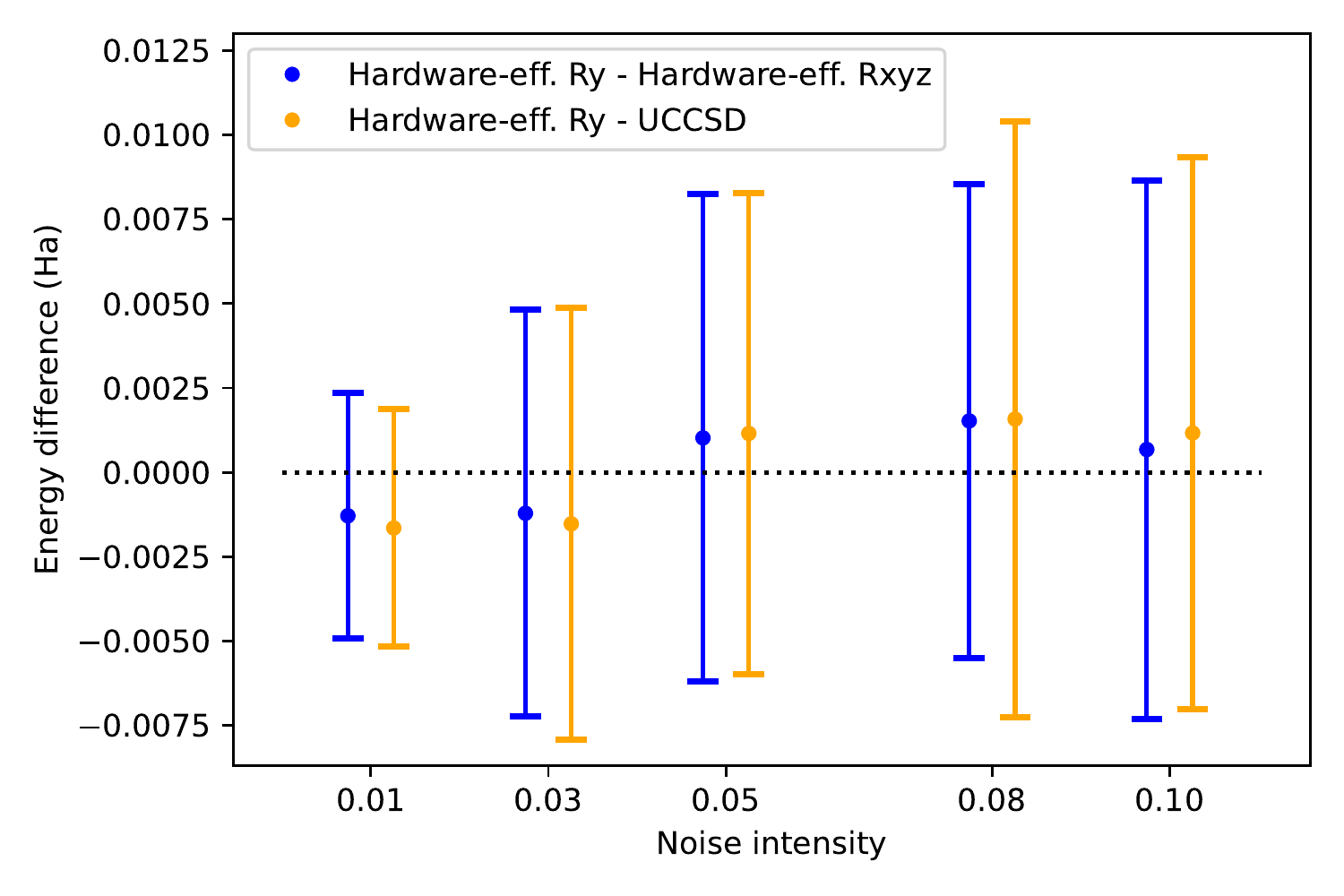}
    \caption{Difference of the mean energy values resulting from simulations with the $R_Y$ and the UCCSD ansatz with respect to simulations with $R_{XYZ}$ for different intensities of readout noise. Note the small scale.}
    \label{fig:RoNoDiff}
\end{figure}

On the other hand, simulations with UCCSD exhibit a different behavior for depolarizing, amplitude and phase damping noise. Compared to the hardware-efficient ansatzes, the effect of noise on the energies is much larger, especially for phase damping noise. The noise-energy relationship follows a characteristic saturation curve, as can be seen in figure~\ref{fig:depHwe}. In particular, the 2-qubit gate depolarizing noise accumulates quickly, leading to a plateau in the energy for noise intensities above $\sim 0.05$.
We tested various 
polynomial and sigmoid functions as potential fit functions and found that the noise-energy relationship is best described by a Gauss error function. 
For readout noise the relationship is modeled with a linear function and the obtained energy values are very similar to the ones from the hardware-efficient ansatzes. Figure \ref{fig:RoNoDiff} shows that the results for the different ansatzes agree well within the uncertainties. 

The independence of the results from the type of rotation used in the hardware-efficient ansatzes and the independence of the effect of readout noise from the ansatz suggest that the energy shift mainly depends on the total accumulated noise. Since readout noise is only applied once before the measurement, it is independent of the number of gates in the circuit and thus the total amount of noise is the same for both the hardware-efficient and UCCSD ansatzes.
The other noise types constitute gate-based errors. Therefore, the
larger effect these noise types have on the results with the UCCSD ansatz may be attributed to the fact that more noise is accumulated in the longer circuit.

\begin{figure}[h!]
    \centering
        \includegraphics[width=0.5\textwidth]{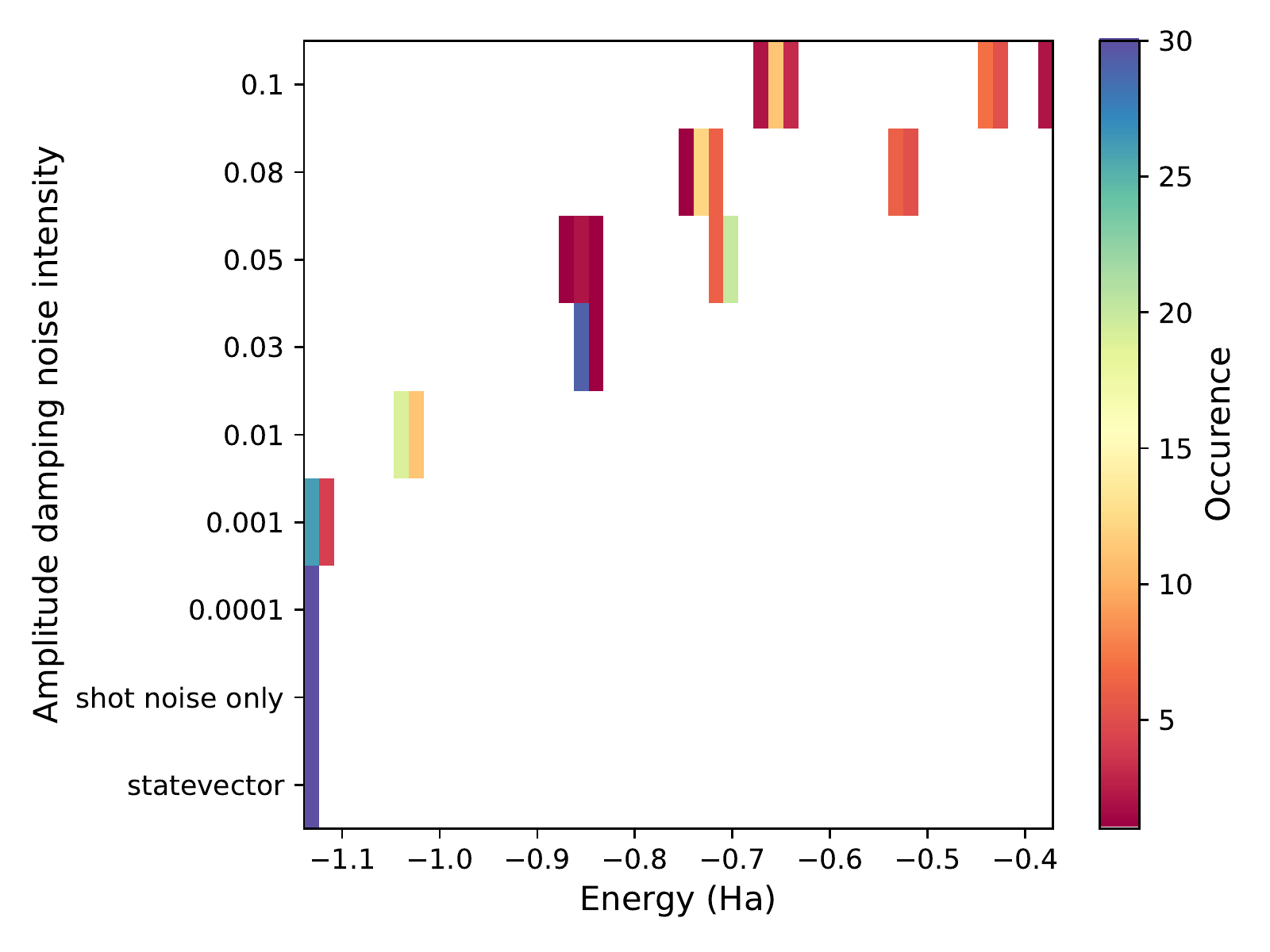}
    \caption{Histograms of the resulting energies of 30 repetitions of VQE for various intensities of amplitude damping noise. The simulations are done with the UCCSD ansatz and 8192 shots.}
    \label{fig:ampSplit}
\end{figure}

With amplitude damping noise in simulations with UCCSD a new effect arises, which is not seen in the other cases and is depicted in figure \ref{fig:ampSplit}: a splitting of the resulting energies into two levels with increasing distance for higher noise intensities. A similar behaviour was also observed in~\cite{fontana_evaluating_2021} and we study this effect in more detail in the next section.

\subsection{Further investigations on the optimization process and solution quality under noise}

\begin{figure}[h!]
    \centering
    \includegraphics[width=0.72\textwidth]{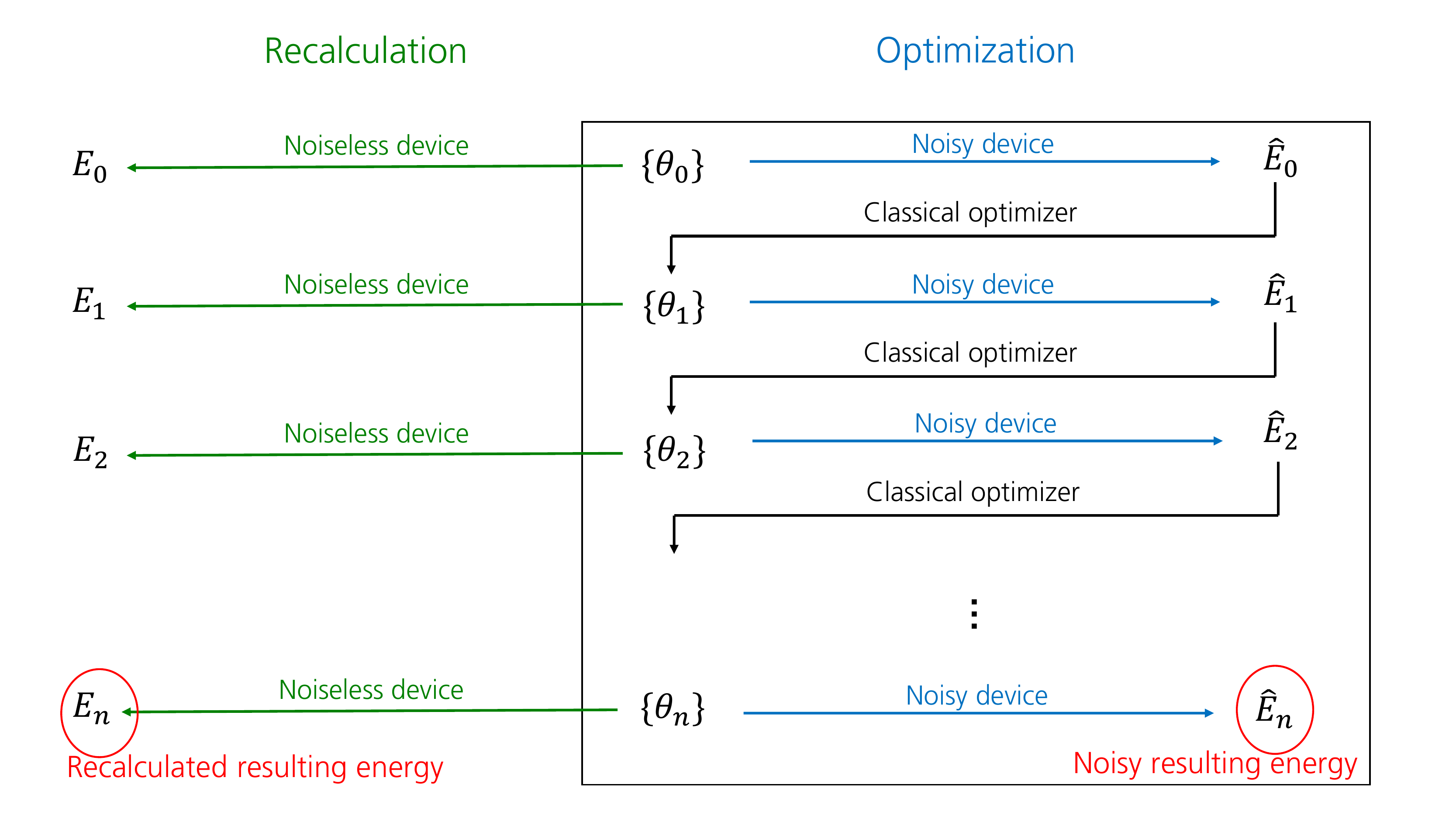}
    \caption{Principle of the recalculation method: After the optimization is finished the energies corresponding to the sets of parameters are recalculated in a noiseless simulation for each iteration. The iteration number is denoted by the subscript index.}
    \label{fig:recalculation}
\end{figure}

In this section we further investigate the noise-induced energy shift and the energy splitting seen with amplitude damping noise.
In these studies we consider not only the found energies but also the parameters found by the optimizer. For this, we use a technique that we call recalculation and adapted from~\cite{mihalikova_cost_2022}. Its principle is illustrated in figure \ref{fig:recalculation}.

\begin{figure}[h!]
    \centering
    \includegraphics[width=0.49\textwidth]{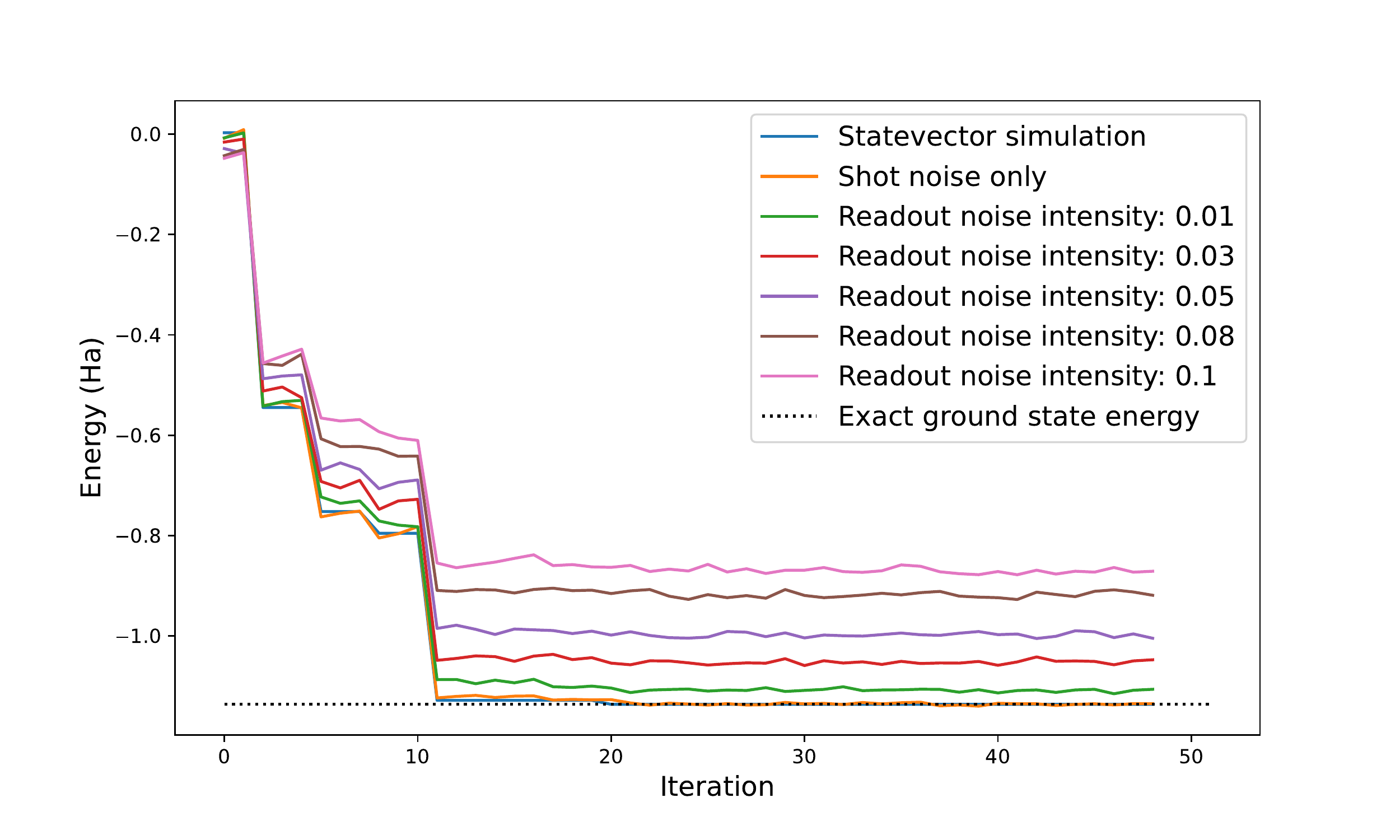}
    \includegraphics[width=0.49\textwidth]{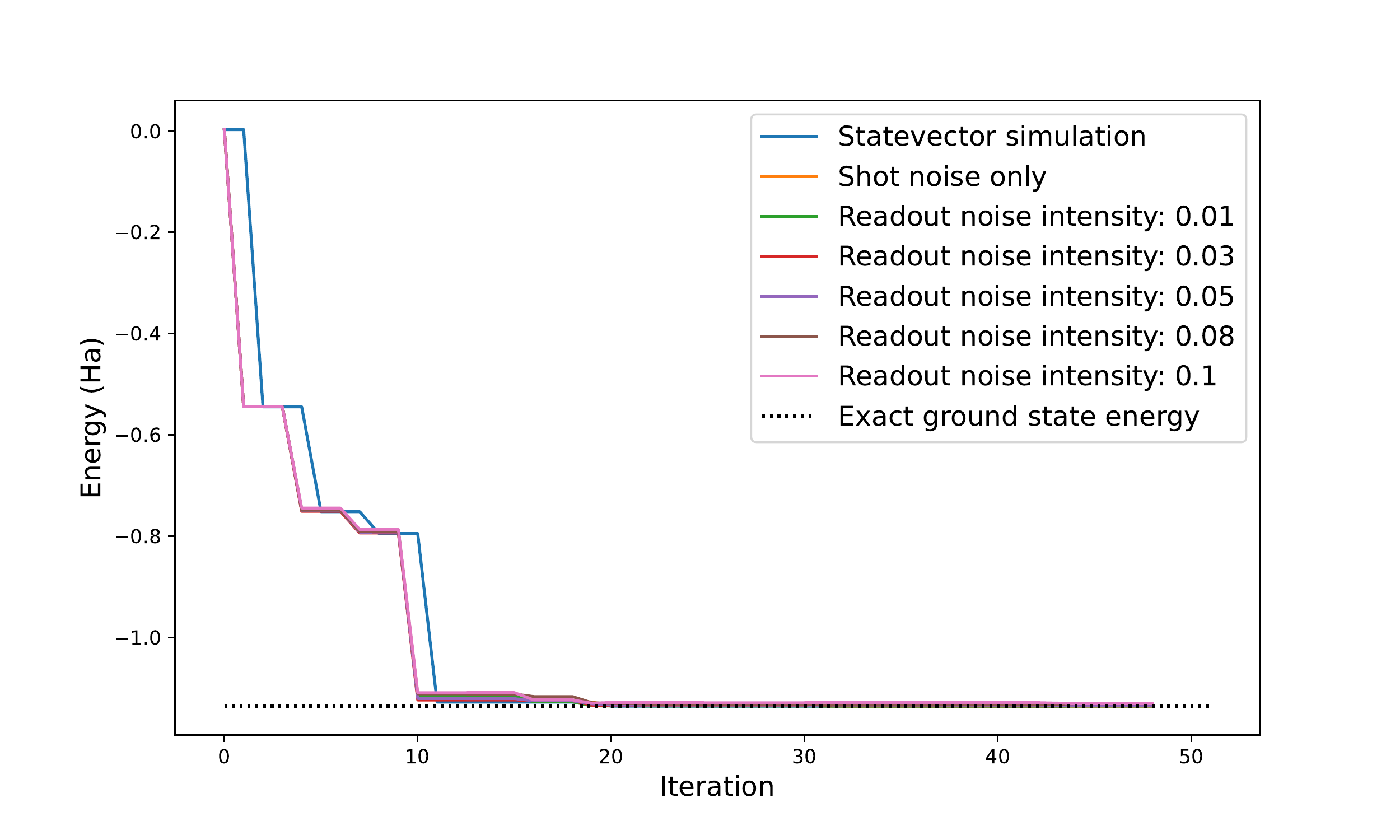}
    \caption{Energy as function of the iteration. Left: calculated with simulations with different levels of readout noise during the optimization, Right: the corresponding recalculated energies per iteration.}
    \label{fig:recalcEvsIt}
\end{figure}

In the noisy VQE, in each optimization iteration the energy corresponding to a set of parameters is calculated by a noisy device. The result of this noisy loss function evaluation is used by the classical optimizer to update the parameters in the next iteration until the optimization is stopped. In the recalculation the loss function is evaluated again after the optimization has been finished at the found sets of parameters using a noiseless simulation. No further optimization is done.

The left plot in figure \ref{fig:recalcEvsIt} shows the energies found in the noisy optimization as a function of the iteration for simulations with the $R_{XYZ}$ ansatz and different intensities of readout noise. The right plot shows the corresponding recalculated energies per iteration. 
While the energies computed in the noisy simulations are shifted with increasing noise intensity, the parameters found by the optimizer yield energies near the exact ground state energy for all cases in the noiseless recalculation. This effect is observed for all kinds of noise.

The absence of the energy shift in the recalculated energies means that the NFT optimizer finds parameters which are optimal in the noiseless case. This suggests that the shift is introduced by the noisy evaluation of the loss function in each single iteration, not the optimization itself. Thus, the noise has the effect of increasing the value of the loss function at the point of the minimum of the noiseless loss function. It further implies that for the accuracy of the VQE the noise intensity in the last iteration is critical, as described by \cite{mihalikova_cost_2022} for shot noise. This matches our observation that VQE simulations with high noise levels during the optimization, but low noise in its last step, give similar accuracies as simulations with low noise levels in the whole optimization process.This insight could be used for future decisions on the choice of the hardware device. Computational resources
could be saved with little loss of accuracy by performing the optimization on a device with worse noise properties and a relative low number of shots and afterwards recalculating the energy corresponding to the found optimal parameters on a device with better noise properties and a high number of shots. 

\begin{figure}[h!]
    \centering
        \includegraphics[width=0.49\textwidth]{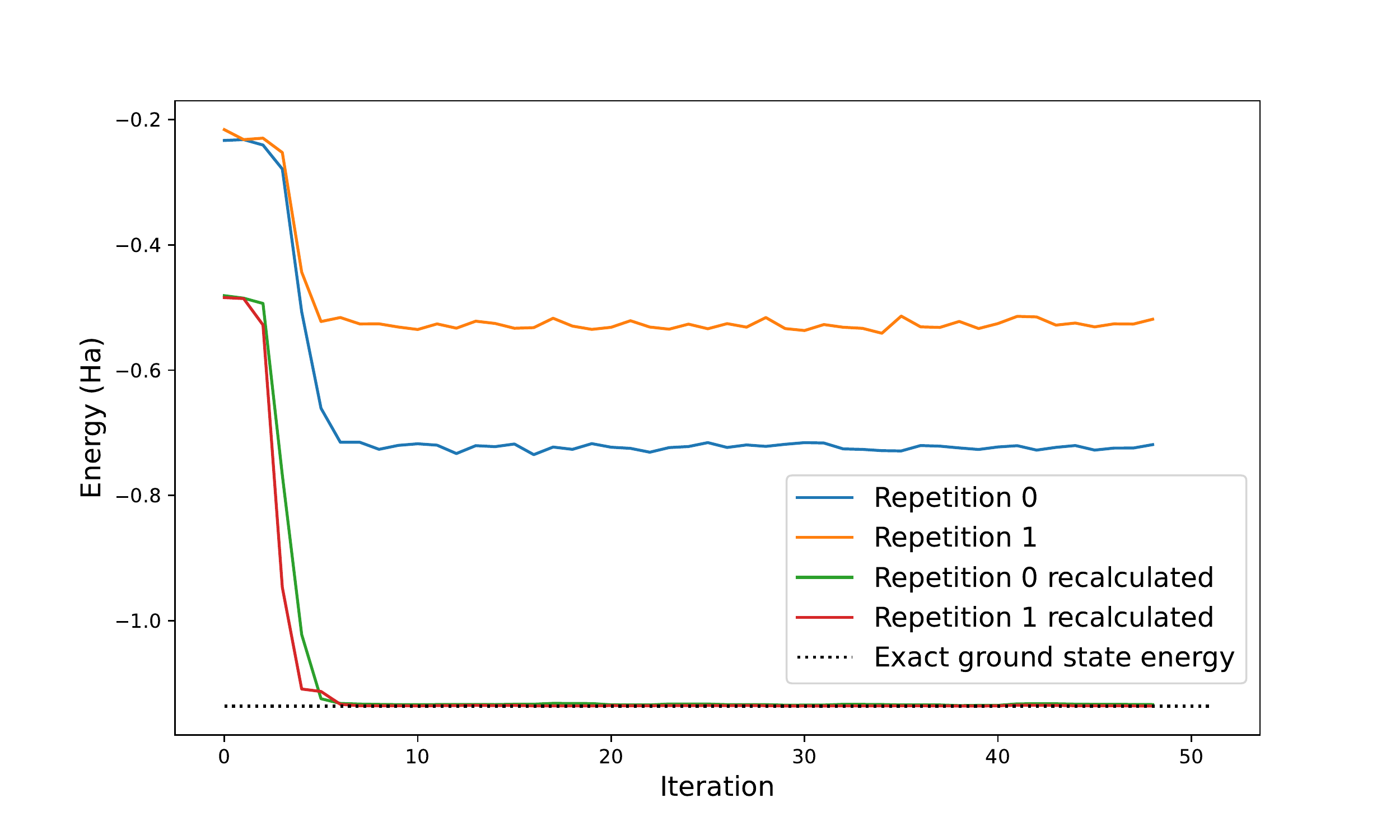}
        \includegraphics[width=0.49\textwidth]{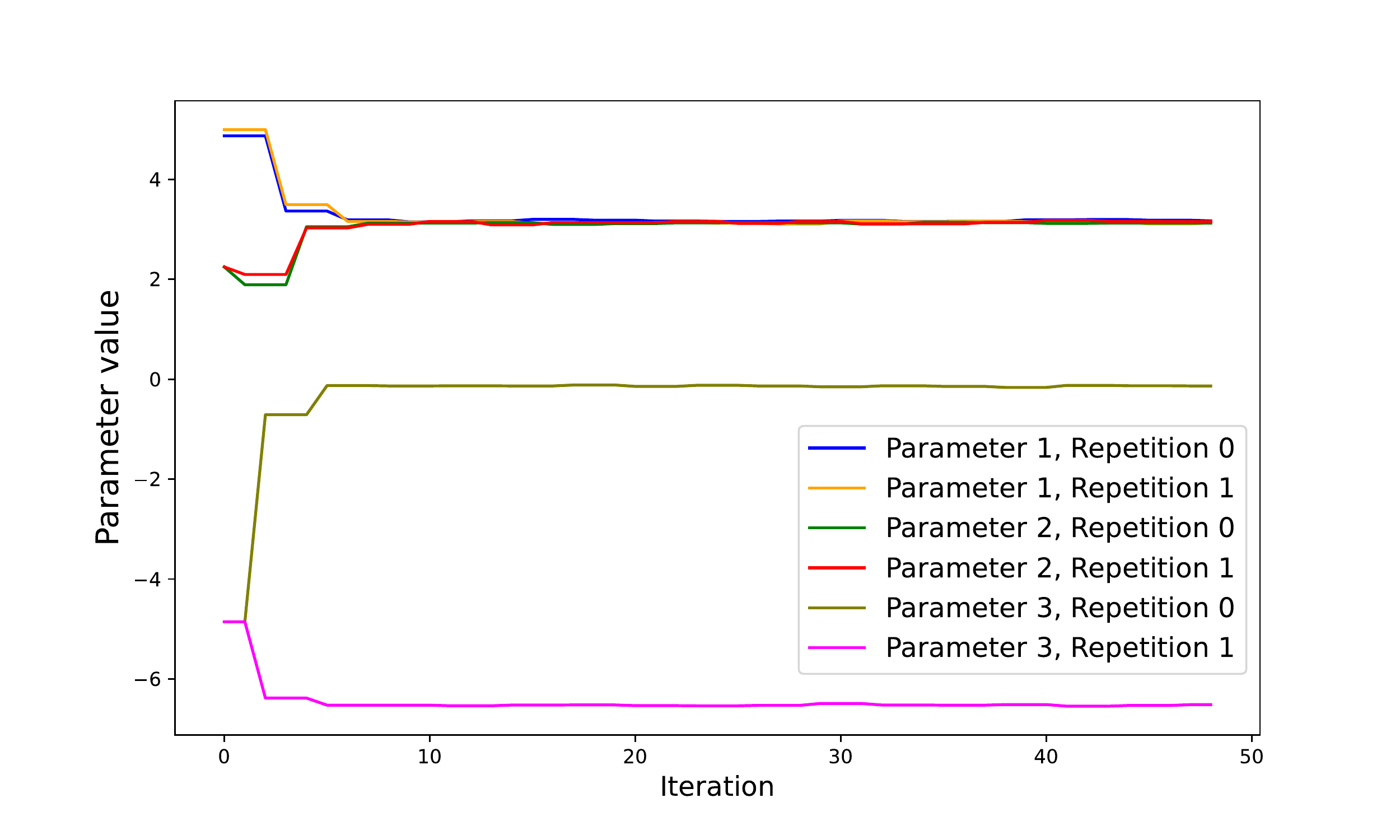}
    \caption{Splitting of the resulting energies for the UCCSD ansatz using amplitude damping noise with an intensity of 0.08. Left: Energy in the optimization process as function of the iteration for two repetitions and their corresponding recalculated values. The two repetitions are exemplary for runs which give a resulting energy on one of the two levels. Right: Parameters in the process of optimization as function of the iteration for the two mentioned repetitions.}
    \label{fig:ampSplit_recalc}
\end{figure}

Figure~\ref{fig:ampSplit_recalc} depicts more details of the energy splitting for amplitude noise with the UCCSD ansatz.
The left plot shows that the splitting arises from early stages in the optimization on. One can also see that the parameters in the iteration, where the optimization saturates, correspond to energies near the exact ground state energy when recalculated with a statevector simulation. 
The obtained parameters are shown in the right plot of figure~\ref{fig:ampSplit_recalc}. These differ between each other by multiples of 2$\pi$. As these parameters are applied in rotation gates, this means that they are effectively equivalent, leading to the same energies in the noiseless case. The energy splitting can be explained by a symmetry breaking in the loss landscape, which is described in more depth in \cite{fontana_evaluating_2021}. A degeneracy of sets of parameters, corresponding to the same energy in the noiseless case, can be lifted in presence of noise.

%% file: conclusion.tex
\label{sec:conclusion}

In this work, we systematically studied the impact of noise on the performance of the VQE algorithm using as example problem the calculation of the ground state energy of the hydrogen molecule.  We first tested the behavior of different classical optimizers and found NFT to be best suited for the subsequent studies, as it can deal well with noise. Next, we investigated the effect of different types of noise in more detail. We varied the noise intensity of each type separately around the region of typical noise levels of IBM Quantum Falcon version 1 processors~\cite{ibmq}. We find that a decreasing number of shots broadens the distribution of resulting energies, while its mean value is unaffected. Other types of noise introduce a shift of the energies towards higher values. We find mathematical models for the relationship between the mean value of energies and the noise intensity. This relationship depends on the type of noise as well as on the ansatz used for the simulation. 
Furthermore, we observe that the sets of parameters found in the noisy optimization correspond to energies near the exact ground state energy when being recalculated with a noiseless device. This implies that the energy shift is not introduced by the optimization, but by changes in the loss landscape at the positions of minima due to the noise. Further investigations on the shape of the loss landscapes will lead to a deeper understanding of this effect.

A topic of future research is to evaluate our results in comparisons with experiments done on quantum processors. Furthermore, one also needs to study other noise effects in more detail. For example, cross talk between qubits may play an important role for the results calculated on real hardware. Furthermore, since our studies focused on IBM superconducting hardware, it is necessary to understand these effects on other devices. On the way to future applications of the VQE algorithm, it will also be necessary to evaluate the impact of noise for larger molecules as well.

%% file: appendix.tex
\graphicspath{{./imagesAppendix/}}
\section{Experimental setup}
\subsection{Hamiltonian} \label{appendix:hamiltonian}

The Hamiltonian is obtained by using the pennylane qchem package \cite{bergholm_pennylane_2018}. The full Hamiltonian is 
\begin{equation} \label{H2HamAppendix}
    \begin{aligned}
        H &=(-0.04207254303152995) I_0
        + (0.17771358191549907) Z_0
        + (0.17771358191549919) Z_1 \\ 
        &+ (-0.2427450172749822) \left(Z_2 + Z_3\right) \\ 
        &+ (0.12293330460167415) \left(Z_0 Z_2+  Z_1 Z_3 \right)
        + (0.16768338881432715) \left(Z_0 Z_3+  Z_1 Z_2\right)\\ 
        &+ (0.17059759240560826) Z_0 Z_1
        + (0.17627661476093917) Z_2 Z_3 \\ 
        &+ (0.04475008421265302) \left( Y_0 X_1 X_2 Y_3+ X_0 Y_1 Y_2 X_3 - Y_0 Y_1 X_2 X_3 - X_0 X_1 Y_2 Y_3\right),
    \end{aligned}
    \end{equation}
where $X$, $Y$ and $Z$ denote the corresponding Pauli operators and their index the qubit it acts on.
\subsection{Ansatzes}
\label{app:ansatzes}
\begin{figure}[h!]
\centering
\includegraphics[width=0.4\textwidth]{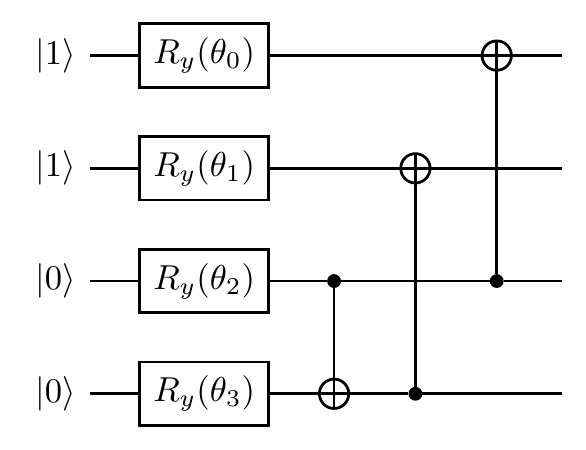}
\caption{The circuit of the hardware-efficient $R_{XYZ}$ ansatz.} \label{fig:ansatzRy}
\end{figure}

\begin{sidewaysfigure}
    \centering
    \includegraphics[height=59pt]{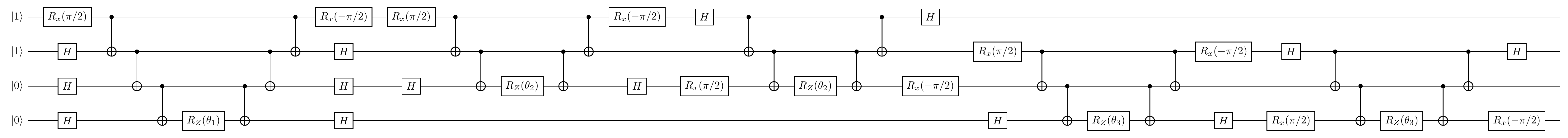}
    \caption{The circuit of the chemically-inspired UCCSD ansatz.}
    \label{fig:ansatzUcc}
\end{sidewaysfigure}

Besides the hardware-efficient $R_{XYZ}$ ansatz, depicted in \ref{fig:ansatzes}, our simulations were done with the hardware-efficient $R_{Y}$ and UCCSD ansatzes which are shown in figures \ref{fig:ansatzRy} and \ref{fig:ansatzUcc}, respectively.


\section{Optimizer studies} \label{appendix:optimizers}
In the following, we briefly describe the different optimizers and the corresponding hyperparameters that must be adjusted before the optimization process.  

\begin{itemize}
    \item Adam, derived from adaptive moment estimation, augments first-order gradient descent by introducing an adaptive learning rate which is based on exponentially decaying averages of gradients of preceding steps. Two hyperparameters ($\beta_1$ and $\beta_2$), the so-called momenta, control the magnitude of influence that previous gradients have on the learning rate. Besides those there are the hyperparameters $\alpha$ corresponding to the non-adapted learning rate and a small additive shift $\epsilon$ to avoid division by zero.  
    \item Nelder-Mead is a gradient-free optimizer which is also commonly used in classical optimization problems. In each iteration this optimizer evaluates the loss function at the vertices of a $N$-dimensional simplex, where $N$ is the number of parameters to be optimized. It updates this simplex based on the results until it converges at the minimum of the loss function. 
    The size of the initial simplex is the only hyperparameter to adjust by the user.
    In some cases it can be beneficial to overcome local minima by restarting the optimization around the previously found position, but with a larger simplex after insufficient convergence. 
    \item SPSA is an approximate gradient descent method where the gradient is estimated by the central difference method. The direction from the current best point to the point of the next loss function evaluation is chosen randomly. Its two hyperparameters are the finite distance $c$ and the learning rate $a$. Due to its stochastic nature, the SPSA optimizer is supposed to deal well with noise.
    \item Bayesian optimization uses Gaussian process regression to create a predictive model of the loss function which is based on the previous evaluations of it. In each step the current model is used to build an acquisition function. The maxima of the function determine the positions of the evaluations in the next step. The bounds of the search space and the number of points to evaluate for the inital model are adjustable hyperparameters. 
    \item The NFT algorithm iteratively picks only one parameter at a time and uses the fact that the loss function in dependence of a single parameter is described by a sine curve to optimize this parameter. This method needs two to three loss function evaluations per iteration, which depends on the single hyperparameter it has: Every x-th iteration a third, additional loss function evaluation is done at the position of the current minimum to prevent gradually enlarging errors, where x is the hyperparameter. Additionally, we introduced a second hyperparameter which determines in which order the parameters are chosen to be optimized: either in an 'ordered' way, where the succession is fixed and periodically run through, or the next parameter to be optimized is chosen randomly from a set of parameters, from which the picked one is removed afterwards to avoid that one parameter is optimized more often than others. When this set is empty it is filled with all parameters again.
\end{itemize}
For each optimizer additionally either the number of the total iterations or a convergence criterion must be specified. First, different configurations have been tested for each optimizer and optimized with spot tests. The best of the configurations for each optimizer have then been compared as shown in figure \ref{fig:allOptimizers}.

\begin{figure}[h]
    \centering
    \includegraphics[width=0.45\textwidth]{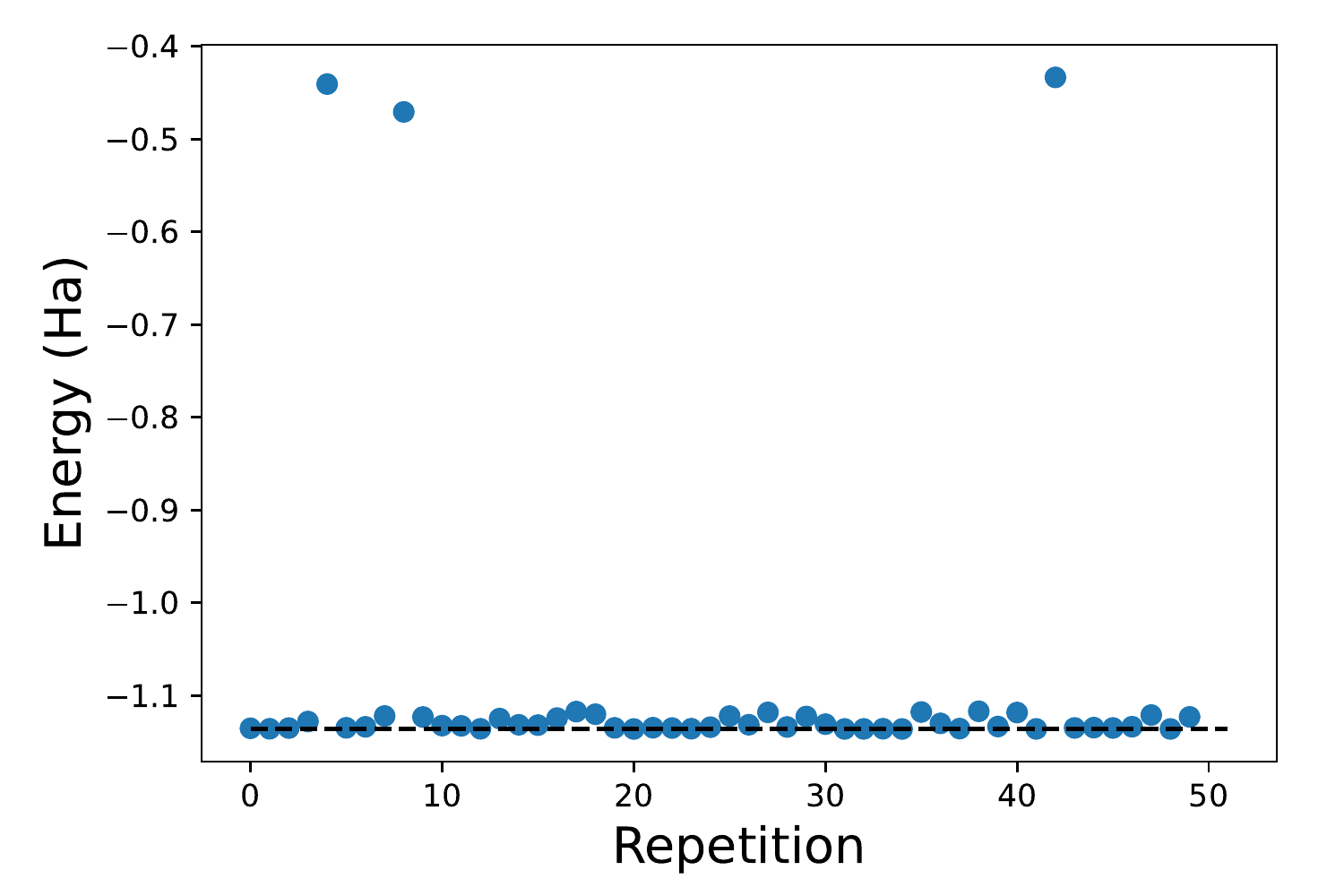}
    \includegraphics[width=0.45\textwidth]{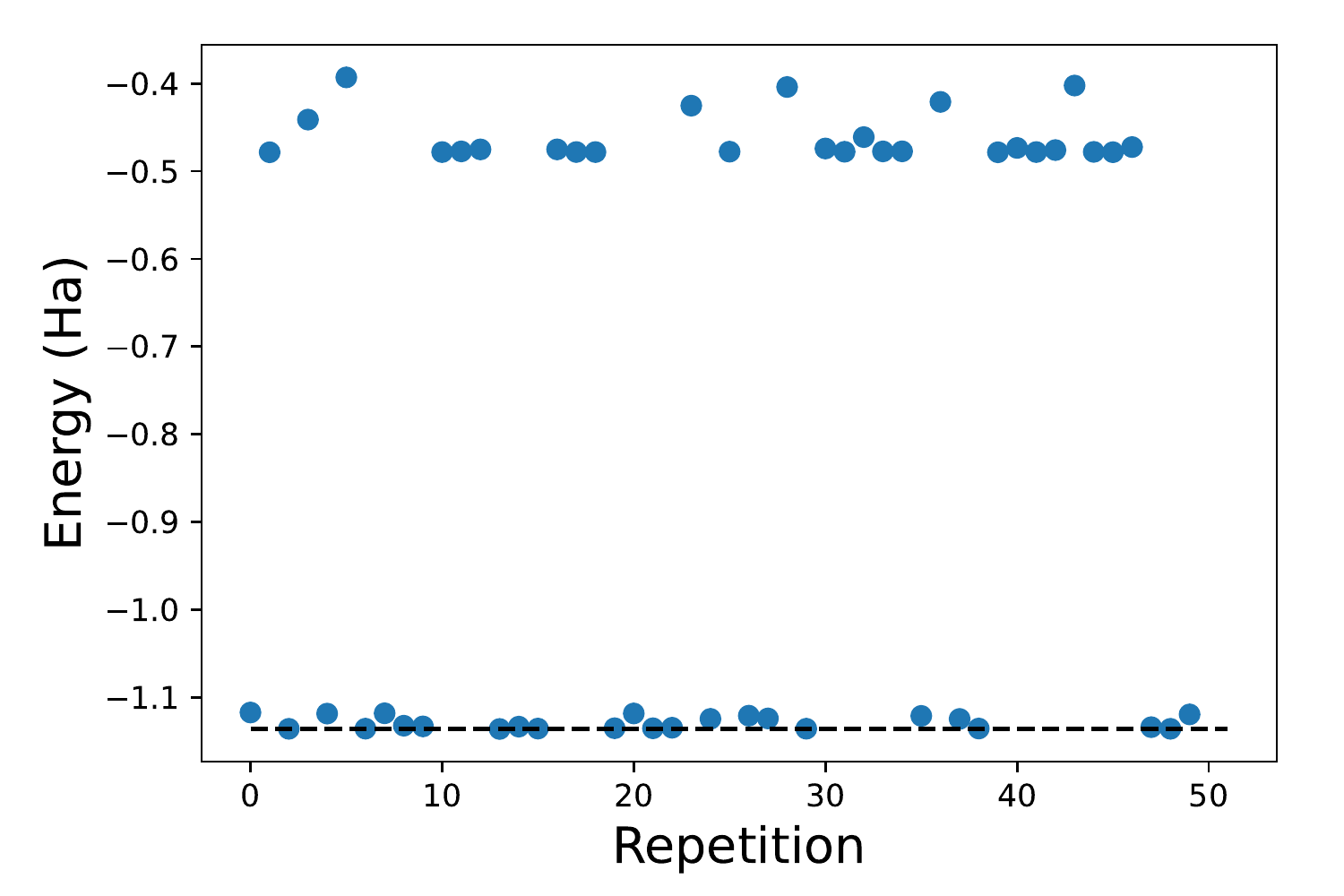}
    \caption{The results of VQE for the H2 molecule for 50 repetitions with random sets of initial parameters for optimization with SPSAreopt (left) and original SPSA (right).}
    \label{fig:spsaroptComp}
\end{figure}

\paragraph{SPSAreopt}
For SPSA we introduce a modification to the original algorithm, which we refer to as SPSAreopt in the following.
The performance of the original SPSA algorithm can be improved by introducing a two-step process with a coarse search followed by a finer one:
 With hyperparameters near those recommended by general guidelines \cite{james_c_spall_implementation_1998} the optimization with SPSA frequently converges in local traps. This can be overcome by choosing much larger hyperparameters, but the global minimum is eventually departed again before the fixed number of iterations has been performed. Adding a convergence criterion to SPSA and optimizing with hyperparameters following the guidelines only after the criterion was met in a optimization with large hyperparameters enables finding good results for several sets of initial parameters which lead to convergence in local traps otherwise.
Figure \ref{fig:spsaroptComp} shows the improved performance of SPSAreopt with respect to the original SPSA.

